# Real-Time Health Analytics Using Ontology-Driven Complex Event Processing and LLM Reasoning: A Tuberculosis Case Study

Ritesh Chandra*, Sonali Agarwal, and Navjot Singh
Indian Institute of Information Technology Allahabad, Prayagraj, India
rsi2022001@iiita.ac.in, sonali@iiita.ac.in, navjot@iiita.ac.in

## Abstract

Timely detection of critical health conditions remains a major challenge in public health analytics, especially in Big Data environments characterized by high volume, rapid velocity, and diverse variety of clinical data. This study presents an ontology-driven real-time analytics framework that integrates Complex Event Processing (CEP) and Large Language Models (LLMs) to enable intelligent health event detection and semantic reasoning over heterogeneous, high-velocity health data streams. The architecture leverages the Basic Formal Ontology (BFO) and Semantic Web Rule Language (SWRL) to model diagnostic rules and domain knowledge. Patient data is ingested and processed using Apache Kafka and Spark Streaming, where CEP engines detect clinically significant event patterns. LLMs support adaptive reasoning, event interpretation, and ontology refinement. Clinical information is semantically structured as Resource Description Framework (RDF) triples in GraphDB, enabling SPARQL-based querying and knowledge-driven decision support. The framework is evaluated using a dataset of 1,000 Tuberculosis (TB) patients as a use case, demonstrating low-latency event detection, scalable reasoning, and high model performance (in terms of precision, recall, and F1-score). These results validate the system's potential for generalizable, real-time health analytics in complex Big Data scenarios.

## 1. Introduction

TB remains one of the most critical global public health challenges. According to the World Health Organization (WHO)[1], in 2024, 193 countries representing over 99% of the global population reported TB-related health data. In 2023, TB once again became the world's leading infectious disease killer, surpassing COVID-19[2], with approximately 1.5 million deaths globally, including around 360,000 among HIV-positive patients[3]. India remains disproportionately burdened, contributing over one-fourth of the global TB cases, with nearly 28 lakh new infections and approximately 3.2 lakh deaths reported in 2023, as highlighted in the Global TB Report 2024 [1].

Even though India has national programs like the National Tuberculosis Elimination Programme (NTEP) and the Revised National Tuberculosis Control Programme (RNTCP), there are still many problems in diagnosing and treating TB, especially in rural and poor areas [2]. Many patients face delays in getting diagnosed because there are not enough specialists, and health records are often scattered across different systems. People in remote areas may also lack sufficient awareness about TB symptoms and treatment options. On top of this, the healthcare system must handle large volumes of data coming from hospitals, laboratories, community health workers, and field visits. These datasets arrive at high velocity, often in real time, and exhibit significant variety, ranging from structured patient records to unstructured reports and observational notes. These characteristics of volume, velocity, and variety, the "3Vs" of Big Data, make it increasingly difficult for healthcare professionals to make quick and accurate decisions.

---

1 https://www.who.int/teams/global-programme-on-tuberculosis-and-lung-health/tb-reports/global-tuberculosis-report-2024
2 https://www.who.int/news/item/29-10-2024-tuberculosis-resurges-as-top-infectious-disease-killer
3 https://www.who.int/publications/i/item/9789241564809

In this context, ontology-based semantic modeling offers a powerful approach to address the challenges of healthcare big data environments. Ontologies enable structured, machine-interpretable representation of clinical knowledge, supporting semantic harmonization of heterogeneous datasets and facilitating knowledge-driven diagnostic reasoning. When integrated within scalable big data architectures, ontologies not only ensure data interoperability but also support real-time diagnostic inference by connecting diverse clinical inputs to formally defined concepts, attributes, and relationships [3].

To address these challenges, this study proposes a real-time, explainable DSS for TB diagnosis and management, combining ontology-driven knowledge representation with big data stream processing and explainable AI techniques. Designed for mobile and web-based platforms, the system aims to assist healthcare professionals across urban hospitals and remote field settings by promoting early detection, reducing specialist dependency, and lowering healthcare costs through guided, accurate clinical recommendations. The core components of the proposed DSS include:

1. A standardized TB Ontology, developed using the Ontology Web Language (OWL) and aligned with NTEP, RNTCP, and WHO clinical guidelines, to represent structured domain knowledge.
2. SWRL rules embedded within the ontology support automated diagnostic reasoning, clinical event detection, and generate LLM-based explanations that make precautionary suggestions clearer and easier to understand for everyday users.
3. Real-time data ingestion and processing using Apache Kafka, Apache Spark, and the Siddhi CEP engine for handling continuous patient data streams.
4. A semantic data system powered by GraphDB stores patient data and inferred clinical insights as RDF triples, allowing efficient and structured querying using SPARQL.
5. LLMs are used to provide clear, human-readable clinical explanations and help continuously update and improve the ontology by analyzing new clinical guidelines and policies.

By embedding the ontology as a semantic layer within the big data architecture, the system facilitates real-time processing of heterogeneous clinical data streams. Ontology-driven reasoning enables semantic enrichment and diagnostic inference, while GraphDB ensures scalable, structured knowledge retrieval during clinical workflows [4][5]. The LLM component further enhances clinical transparency and ensures that the knowledge base evolves to reflect emerging medical knowledge [6]. This study addresses key technical and clinical challenges:

1. Ontology-based integration of heterogeneous clinical data sources, ensuring semantic interoperability and unified diagnostic reasoning.
2. Knowledge-driven decision support using SWRL-based semantic reasoning to improve diagnostic accuracy.

3. Real-time event detection through CEP combined with ontology rules, enabling early identification of critical diagnostic indicators.
4. Scalable semantic querying and retrieval using GraphDB and SPARQL to deliver rapid, context-specific clinical recommendations.
5. Continuous ontology evolution, supported by LLMs analyzing new clinical knowledge for automated updates and refinements.

The structure of the remaining sections is as follows: Section II outlines the background analysis along with a review of related literature. Section III describes the proposed system and explains its functional architecture. Section IV provides insights into the experimental setup and showcases the results. Section V concludes the research and highlights avenues for future exploration. Additionally, Table 1 presents the list of abbreviations used throughout the paper.

Table 1 Abbreviations

| **Abbreviation** | **Full Form** | **Abbreviation** | **Full Form** |
|---|---|---|---|
| Tuberculosis | TB | Semantic Web Rule Language | SWRL |
| Complex Event Processing | CEP | Resource Description Framework | RDF |
| Large Language Models | LLMs | World Health Organization | WHO |
| Basic Formal Ontology | BFO | National Tuberculosis Elimination Programme and | NTEP |
| Ontology Web Language | OWL | Revised National Tuberculosis Control Programme | RNTCP |
| Human immunodeficiency virus | HIV | Bidirectional Encoder Representations from Transformers | BERT |
| Ontology-based Complex Event Processing | OCEP | Natural language processing | NLP |
| Patient Clinical Data | PCD | National Viral Hepatitis Control Program | NVHCP |
| Explainable AI | XAI | Optical Character Recognition | OCR |
| Ontology for TB Surveillance System | O4TBSS | Applied Ontology-Based Data Management | OBDM |
| Comma Separated Values | CSV | Designated Microscopy Centers | DMCs |
| Intermediate Reference Laboratory | IRL | Peripheral Health Institutes | PHIs |
| Non-Governmental Organization | NGO | State Training and Demonstration Centres | STDCs |
| State TB Cell | STC | Centre TB Cell | CTC |
| State Tuberculosis Officer | STO | stands for Community Health Centre | CHCs |
| Medical Officer | MOs | Multipurpose Worker | MPWs |
| Multipurpose Health Supervisor | MPHS | Decision Support System | DSS |
| Ministry of Health and Family Welfare | MoHFW | Directly Observed Treatment, Short-course | DOTS |
| Revised National Tuberculosis Control Programme. | RNTCP | Semantic Web Rule Language. | SWRL |
| Accredited Social Health Activist. | ASHA | Facebook AI Similarity Search | FAISS |
| Attribute Richness | AR | Class Richness | CR |
| Average Population | AP | Chief Medical Officers | CMOs |
| Relationship Richness | RR | Multi Agent System | MAS |
| Electronic Health Record | EHR | Ontology-Based Complex Event Processing | OCEP |
| Semantic Sensor Network | SNN | Sensor, Observation, Sample, and Actuator | SOSA |
| Certified Ophthalmic Technician | CoT | Epidemiology Ontology | EPO |
| Ontology-Based Data Access | OBDA | Fast Healthcare Interoperability Resources | FHIR |

## 2. Background and Related Works

To find the research gaps and understand how the proposed answer fills them, it is important to first look at previous studies in the field and the problems they had, then go into more detail in the sections that follow.

### 2.1 Ontology Approaches in big data analytics

In the era of Big Data, where information is characterized by high volume, velocity, and variety, traditional data management techniques often fall short in extracting meaningful insights from unstructured or semi-structured data. Ontology-based approaches address these challenges by providing a semantic layer that formally defines concepts, relationships, and rules within a domain [7]. This enables semantic interoperability across disparate data sources, facilitating more effective data integration, intelligent querying, and automated reasoning [8]. Ontologies act as shared vocabularies and logical frameworks that bridge the gap between machine-readable and human-understandable knowledge, thus enhancing the quality of data analytics [9] [10]. Moreover, they support context-aware processing, disambiguation, and event correlation, especially in complex domains like healthcare, IoT, and cybersecurity. In Big Data pipelines, ontologies can be integrated with semantic web technologies (like RDF, OWL, and SPARQL) and complex event processing systems to enable scalable and real-time analytics [11][12].

### 2.2 Apache Spark and Kafka

Apache Spark is a unified analytics engine widely used for building real-time big data applications. It accommodates diverse workloads including batch processing, interactive SQL queries, machine learning tasks, and real-time data streaming within a unified framework, eliminating the need for multiple systems and streamlining the development process. Spark's in-memory computing, high speed, scalability, and built-in fault tolerance make it ideal for large-scale analytics. When integrated with Apache Kafka, Spark can consume and process real-time data streams with low latency, making it suitable for applications that demand instant insights [13].

Both Kafka and Spark are inherently fault-tolerant and scalable. Kafka achieves this by replicating topic partitions across multiple brokers, ensuring data durability and high throughput. Spark, in turn, distributes computation across a cluster and uses lineage information to recover lost data. Spark Streaming supports micro-batch and windowed computations, allowing operations such as aggregations over fixed intervals (i.e., every 5 seconds) [14][15]. This combination is widely used in real-time scenarios like fraud detection, IoT monitoring, predictive maintenance, and event-driven analytics, where continuous data processing and responsiveness are critical.

### 2.3 Complex Event Processing

CEP is an advanced method used to analyze and respond to high-speed, continuous streams of data in real-time. It enables the identification of significant patterns, relationships, or anomalies across multiple event sources, often within milliseconds. CEP is essential for applications that demand rapid detection and reaction to critical events such as fraud attempts, system failures, or medical emergencies. Industries like finance, telecommunications, healthcare, transportation, and cybersecurity leverage CEP to make intelligent, time-sensitive decisions [16].

CEP distinguishes between simple and complex patterns. Simple patterns may involve single events meeting specific conditions, while complex patterns reflect relationships or dependencies across multiple events over time. These patterns are defined using specialized CEP query languages (such as SiddhiQL or Esper) that support temporal logic, event sequencing, filtering, and correlation [17]. By continuously evaluating rules against incoming streams, CEP systems enable instant detection and response, supporting use cases like real-time monitoring, alerting, automation, and predictive analytics [18].

### 2.3 Large Language Models

LLMs are advanced AI models trained on massive volumes of textual data to understand, generate, and reason with human language. They use deep learning architectures—primarily transformer-based models like GPT, BERT, or T5—to capture complex language patterns, contextual relationships, and semantic meaning. LLMs can perform a wide range of NLP tasks such as text generation, summarization, translation, question answering, and reasoning without task-specific training [19].

LLMs are increasingly being integrated into real-time analytics and decision-making systems, including domains like healthcare, finance, cybersecurity, and legal analytics [20][21]. When combined with data streams or event-driven architectures (i.e., Kafka + Spark + CEP), LLMs enhance the ability to interpret unstructured data (i.e., clinical notes, alerts, logs) and generate human-like insights or recommendations. Their ability to understand complex queries, infer missing context, and provide knowledge-rich outputs makes them highly valuable in applications requiring intelligent language understanding and adaptive responses.

### 2.4 Related Work

Recent research introduced the Ontology-based Complex Event Processing (OCEP) framework to address semantic heterogeneity and context-awareness challenges in traditional CEP systems. OCEP integrates ontologies with RDF and SPARQL for semantic reasoning over real-time event streams, ensuring interoperability across diverse data sources. Implemented using Hadoop and Kafka, the framework supports scalable storage and real-time event execution. A healthcare case study using IoT sensor data demonstrated 85% accuracy in early illness detection, showcasing OCEP's potential for intelligent, real-time decision support in Big Data environments [22].

Mavridis et al. [23] propose a methodology that leverages LLMs to enhance medical ontology mapping for RDF knowledge graph construction. It evaluates six systems—GPT-4o, Claude 3.5, Gemini 1.5, Llama 3.3, DeepSeek R1, and BERTMap—using a novel framework combining precision, recall, F1-score, and semantic accuracy. The approach integrates LLM-based semantic mapping with BioBERT embeddings and ChromaDB for efficient concept retrieval. Experiments on 108 medical terms show GPT-4o achieving the highest performance with 93.75% precision and a 96.26% F1-score, demonstrating LLMs' effectiveness in improving semantic interoperability in medical data.

Chandra et al. [24] presents a comprehensive DSS for liver disease diagnosis and treatment using ontologies like BFO and Patient Clinical Data (PCD), aligned with NVHCP guidelines for accuracy. Detection rules derived from a decision tree are encoded in SWRL and executed using Pellet and Drools within Protégé. Apache Jena performs batch processing, while SPARQL enables direct querying of detected events. Based on 615 patient records, the system predicts liver disease types and offers tailored suggestions. It integrates OCR for test result extraction and Explainable AI (XAI) for transparent, API-based recommendations, forming an intelligent and automated diagnostic model.

Epidemiological surveillance systems often rely on relational databases and SQL for data storage and analysis, but they lack semantic relationships needed for automated reasoning. To address this, Jiomekong et al. [25] developed an ontology-based approach. Since no existing ontology for TB surveillance was available, the study developed the Ontology for TB Surveillance System (O4TBSS) in collaboration with an epidemiologist. The ontology currently includes 807 classes, 117 object properties, and 19 data properties, enabling enhanced semantic analysis and reasoning capabilities.

Table 2. Comparison of related works

| References | Research focus | Methodology | Results |
|---|---|---|---|
| Kokash et al. [26] | Ontology- and LLM-based data harmonization for federated learning in healthcare | Two-step data alignment using vector embeddings and LLM-supported ontology mapping; deployed in Vantage6 and Brane FL frameworks. | Achieved semantic interoperability across EHRs; effective integration in real-world FL project with Dutch hospitals. |
| Guarnier et al. [27] | Ontology-driven conceptual model for tuberculosis diagnosis | SABiO methodology with UFO foundational ontology to develop OntoTB; validated using expert reviews. | Produced structured ontology (OntoTB) with graphical representations and competence questions; facilitates future data collection and analysis tools. |
| Rahmani et al. [16] | IoT-aware data analysis in healthcare using CEP | Designed 3-layer IoT architecture (context, event, service); CEP applied in event layer; real-time wireless body area network simulation. | Improved real-time data analysis, reliability, and healthcare service quality through CEP-based processing. |
| Sen et al. [11] | Ontology-based NoSQL schema design for semi-structured/unstructured healthcare data | Designed MongoDB schema using healthcare ontology and query patterns; evaluated performance against relational model. | Ontology-based design outperformed traditional RDBMS; showed faster query response and better adaptability for Big Health Data. |
| Kumar et al [17]. | CEP-Enabled Fuzzy Rule-Based Model for Predicting Cardiovascular Conditions | Integrated Apache Kafka, Spark, Siddhi CEP, and fuzzy logic; rules defined per WHO/clinical parameters. | Real-time model classified synthetic data (1000 samples) into 5 risk categories; validated system's decision support accuracy. |
| Croce et al. [12] | Ontology-based data preparation for healthcare analytics | Applied Ontology-Based Data Management (OBDM) on 13 years of diabetes EMR data; modeled, cleaned, and integrated datasets. | Enabled semantic integration and reuse across research tasks; improved data reliability and interoperability for analytics. |

Previous research has largely concentrated on ontology-driven modeling, complex event processing, or AI-based diagnostics as separate efforts. A structured overview of these approaches is presented in Table 2 [11], [12], [16], [17], [26], [27]. For example, fuzzy rule-based CEP has been applied for cardiovascular disease prediction, ontology-based frameworks have been developed for chronic disease management, and IoT-aware CEP architectures have improved healthcare data analysis. However, these systems are often domain-specific or limited to offline analytics. In contrast, the proposed work combines ontology, CEP, and LLMs into a unified framework for real-time tuberculosis detection within a single platform. This integration enables semantic interoperability, high-throughput event correlation, and contextual reasoning over dynamic and heterogeneous health streams. Notably, incorporating LLMs for domain-aware inference over RDF data enhances decision-making in ways that traditional CEP- or ontology-only systems cannot, thereby addressing critical challenges of scalability, flexibility, and semantic depth for intelligent, real-time disease analytics.

## 3. Methodology

In this section, we present the layered architecture of the proposed intelligent tuberculosis detection system, which integrates Apache Kafka, Apache Spark, and the Siddhi CEP engine. The framework enables real-time processing of clinical event streams, supports semantic reasoning through a domain-specific ontology, and incorporates knowledge-driven inference using LLMs. Together, these components form a unified, scalable, and explainable system for accurate TB detection and analytics in dynamic healthcare environments.

### 3.1 Complete Architecture Model

Figure 1 shows the architecture model of the DSS for a real-time TB detection and decision support framework integrating semantic web technologies, CEP, and LLMs. The system starts with TB patient data that is preprocessed and sent to Kafka and Spark streaming for streaming the data into basic events. These events are handled by the Siddhi CEP engine, which uses rule-based event detection to detect severe conditions and notify them. Such alerts, indicating what type of event occurred (and how serious that event is), are forwarded to the LLMs for semantic interpretation. Concurrently, the preprocessed data is translated into RDF and saved in GraphDB to provide structured storage and reasoning through TB Ontology and SWRL rules. Queries to the ontology using SPARQL access patient views and condition-specific knowledge. We index document embeddings of the clinical text using FAISS for fast similarity search that helps LLMs capture context. LLMs serve a double purpose, analyzing incoming alerts, helping to estimate risk, and proposing ontology updates with new findings, which may be subject to expert validation. Clinical responses such as preventive actions for TB patients are initiated according to the LLM-assisted risk assessment and ontology-based insights. This end-to-end architecture realizes real-time event detection and ontology-based decision-making for TB control.

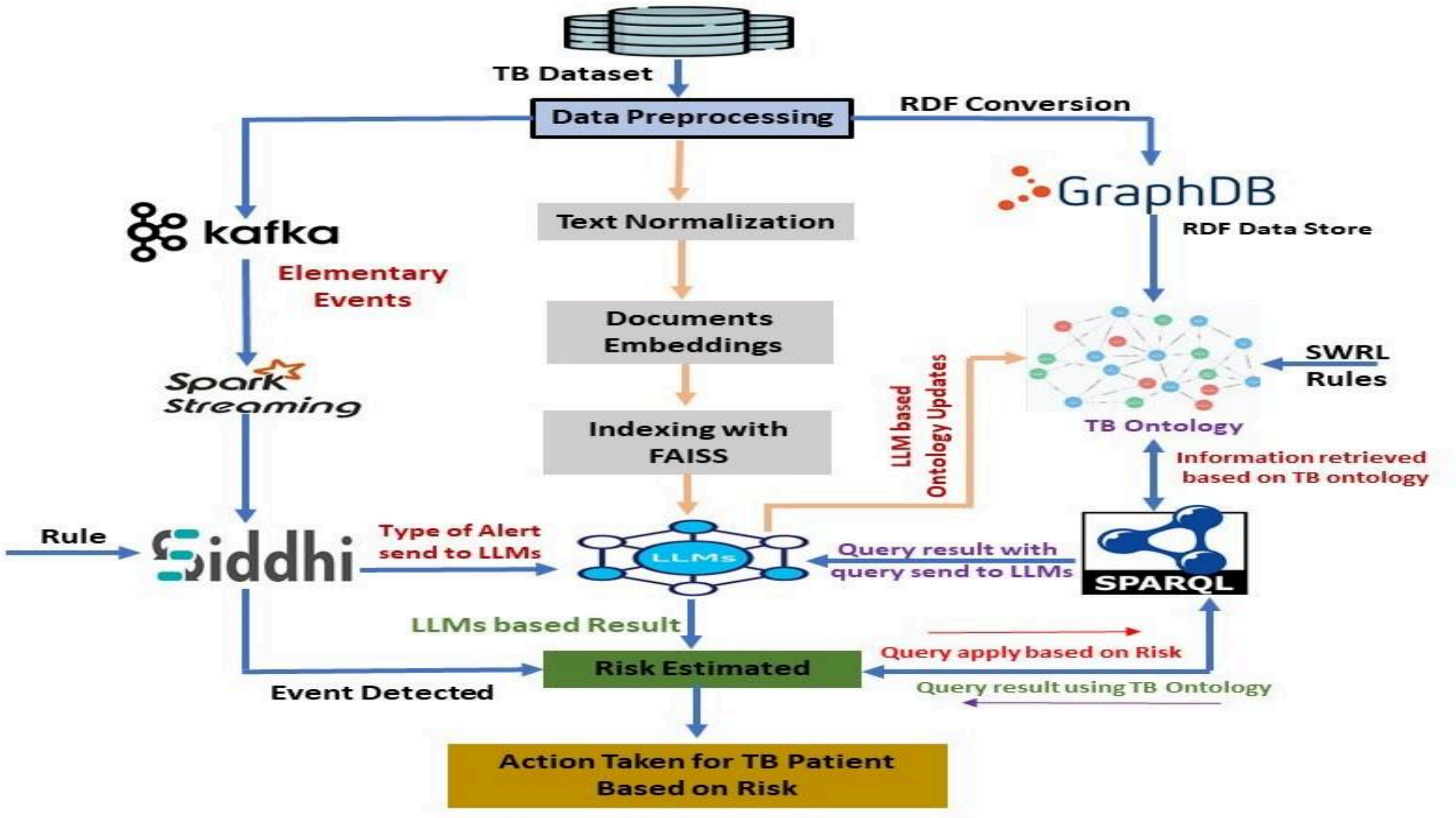

Figure 1. Complete Architecture model of the DSS for TB

### 3.2 Dataset Description

We used the Tuberculosis Symptoms Dataset[4] from Kaggle, which contains 1,000 patient records collected between January 2020 and January 2021. Each record includes patient demographic details such as ID, name, gender, date, and time of observation, along with 13 binary indicators representing tuberculosis-related symptoms. These symptoms include: fever lasting two weeks, coughing blood, sputum mixed with blood, night sweats, chest pain, localized back pain, shortness of breath, weight loss, fatigue, lumps around the armpits and neck, persistent cough and phlegm for two to four weeks, swollen lymph nodes, and loss of appetite.

Since the symptom data was encoded as binary values (0 = No, 1 = Yes), we were able to perform efficient statistical analysis and visualize symptom prevalence across patients. We referred to standard medical guidelines[5] from the WHO, the NTEP, and related protocols to ensure clinical validity. These guidelines informed our system's ontology design, diagnostic rule formulation, and the training of LLMs for semantic reasoning.

### 3.3 Preprocessing

We performed a series of preprocessing steps to prepare the dataset for analysis and real-time processing. First, we removed irrelevant fields such as id, name, and no, which do not contribute to TB detection or decision-making. We then identified and excluded records with missing values in critical fields to maintain data completeness and reliability.

We converted the gender column into a binary format (Male = 1, Female = 0) to facilitate machine processing. We also merged the date and time columns into a single datetime field, from which we extracted the hour and month as additional temporal features. Although all symptom indicators were already binary, we performed logical consistency checks to detect and remove implausible or contradictory entries, such as isolated severe symptoms without any related indicators.

The cleaned and structured dataset was then used as input for our Kafka-based streaming pipeline, semantic enrichment using ontology and RDF triples, and event detection through CEP and LLM reasoning.

### 3.4 RDF Conversion and Store

Figure 2 shows the process of transforming the CSV dataset to RDF format for semantic storage. Following preprocessing, the raw datasets are turned into RDF triples using the RDFLib[6] library. The resulting RDF data is then saved in GraphDB[7], an RDF data store that allows for structured querying and reasoning on the data, then we have integrated with the TB ontology.

---

[4] https://www.kaggle.com/datasets/victorcaelina/tuberculosis-symptoms
[5] https://tbcindia.mohfw.gov.in/guidelines/
[6] https://rdflib.readthedocs.io/en/stable/
[7] https://graphdb.ontotext.com/

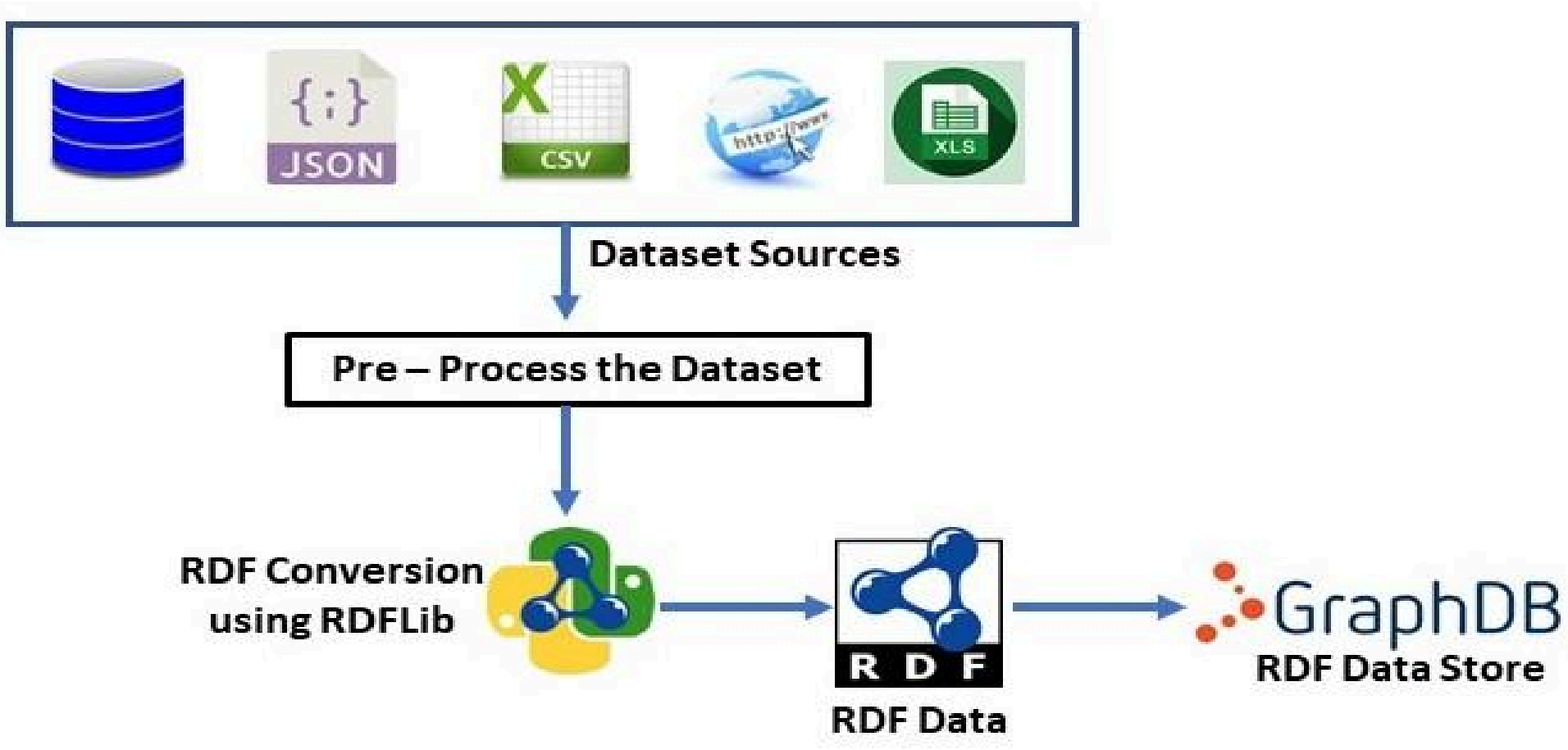

Figure 2. RDF Conversion using RDFLib

### 3.5 TB Ontology Development

Using the RDF data, an ontology is developed based on the concepts and structure defined by the BFO. BFO[8] serves as a top-level ontology framework that facilitates data organization, integration, and retrieval across scientific and other domains. Acting as a foundational model, it provides a standardized approach for representing knowledge. The ontology is a structured form of data representation, where information is organized through classes, attributes, relationships, and individual entities. These entities, which may represent events, conditions, or abstract concepts, are defined using formal semantics to enable human understanding and machine processing. Typically, the ontology is stored and managed within a graph database, where its structure can be visualized as interconnected nodes and relationships [28].

BFO classifies entities into two main categories: continuant classes and occurrent classes. Continuants are entities that exist over time while retaining their identity, even as their attributes or properties may change. These are further divided into three types: independent continuants, specifically dependent continuants, and generically dependent continuants. On the other hand, occurrents refer to entities that unfold, happen, or develop over time. Occurrents are categorized into four subtypes: processes, process boundaries, temporal regions, and spatiotemporal regions, as illustrated in Figure 3. The knowledge graph or ontology based on this classification was created using the Protégé 5.5 software tool.

[8] https://basic-formal-ontology.org/

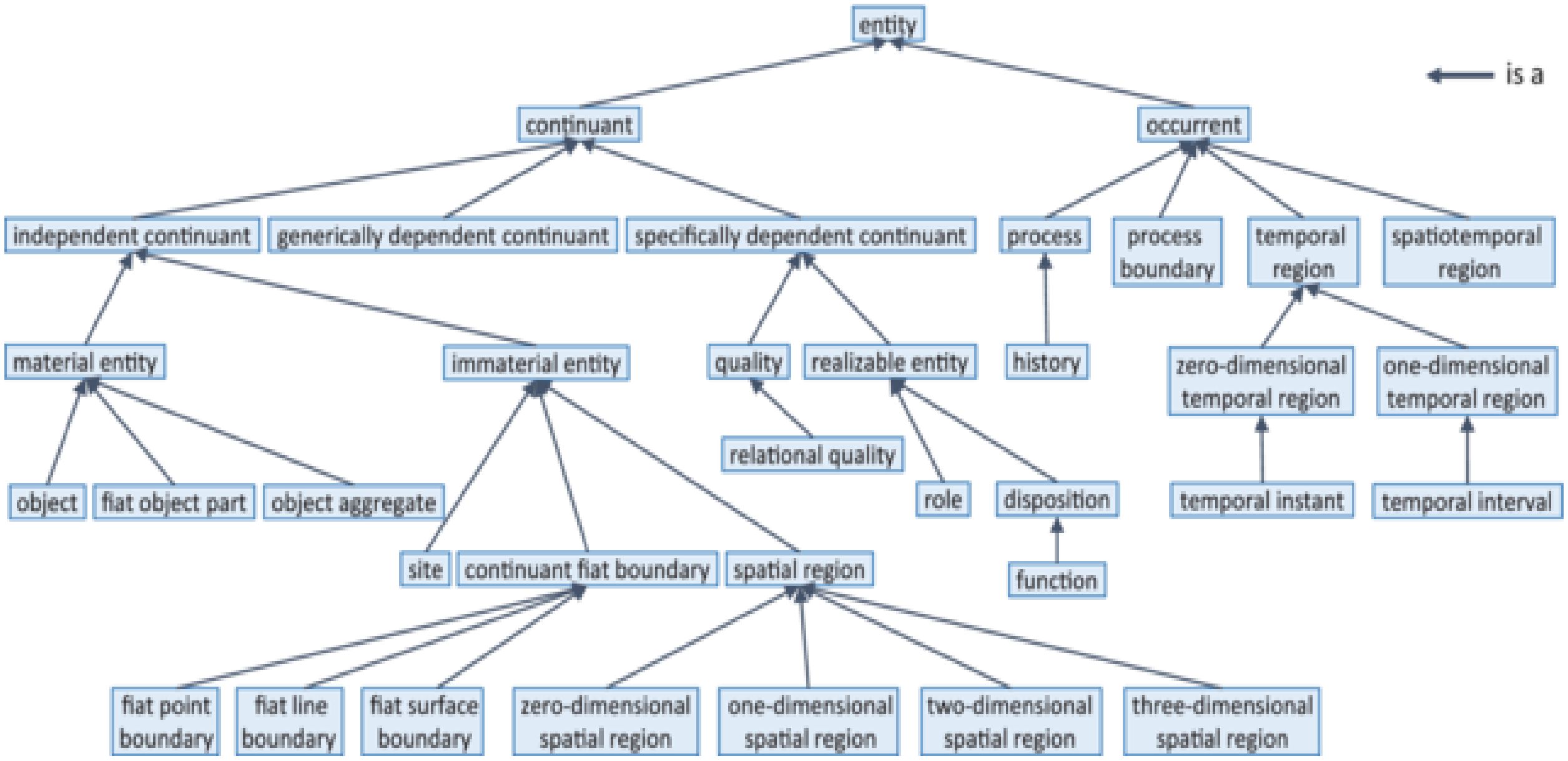

Figure 3. BFO upper-level ontology Hierarchy[9]

### 3.5.1 Core classes of TB ontology

Classes represent categories or types of entities in a knowledge graph. It defines the common characteristics that instances within a class share. Figure 4 presents the classification of Independent Continuants into two categories: immaterial and material entities. Immaterial entities include 'sites' like Designated Microscopy Centers (DMCs), Peripheral Health Institutes (PHIs), and sputum collection centers, which rely on physical structures, as well as 'spatial regions' such as NGOs, WHO offices, hospitals, and medical colleges that operate independently. Material entities include facilities like State Training and Demonstration Centres (STDCs), treated as **'object aggregates'** due to their independent functioning. The State Drug Store, responsible for distributing various medicines, is also categorized as a 'site'. Units like the Training Unit, Monitoring Unit, and Intermediate Reference Laboratory (IRL) are considered part of the STDC as 'object aggregates', although the IRL can function as an independent unit when required. Patients are classified as 'object' entities, as they are central to treatment and may be receiving care for multiple conditions.

---

[9] https://www.iso.org/obp/ui/en/#iso:std:iso-iec:21838:-2:ed-1:v1:en

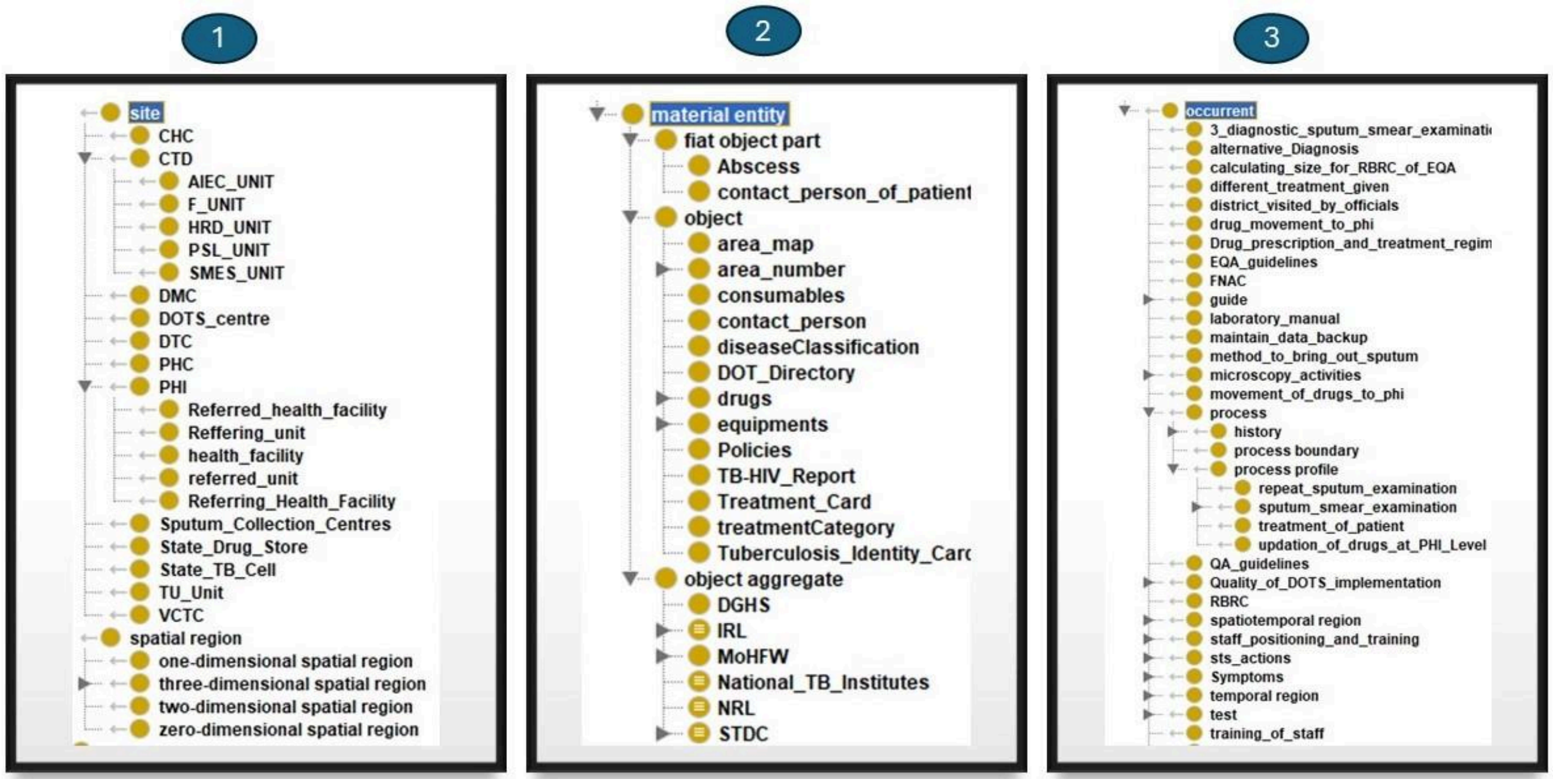

Figure 4. Core classes TB Ontology based on BFO

At the state level, the State TB Cell (STC), led by the State Tuberculosis Officer (STO), manages TB programs following Central TB Division (CTD) guidelines.

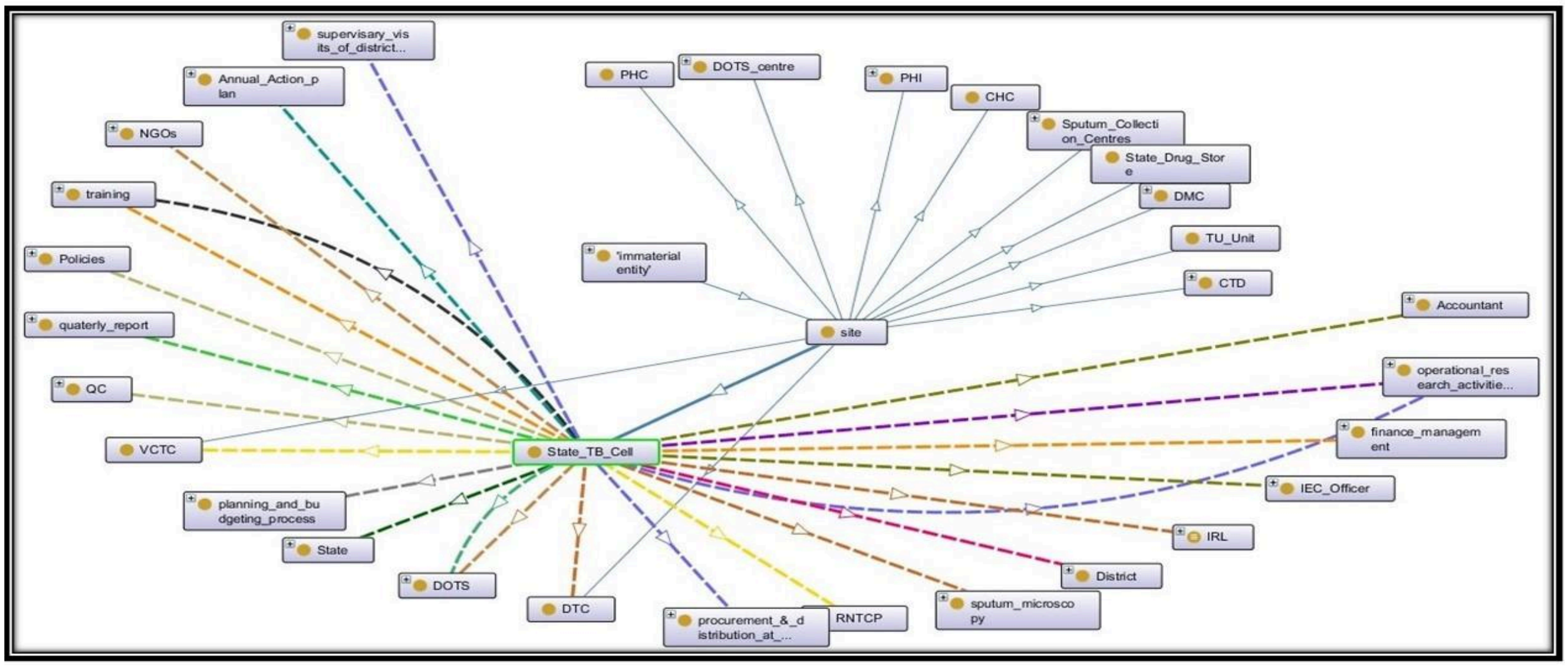

Figure 5. OntoGraf view of STC, its actors, its functions, and its relation with other entities

The STDC works alongside the STC in training and support roles. Staff such as medical officers, pharmacists, and data operators are recognized as ‘roles’, while the STC itself, dependent on the Ministry of Health and Family Welfare (MoHFW) resources, is categorized as a ‘site’ shown in Figure 5 in OntoGraf View.

Peripheral Health Institutions (PHIs), including PHCs, CHCs, referral hospitals, TB hospitals, and medical colleges, function as treatment centers and are also modeled as 'sites', especially those under government control and dependent on MoHFW and STDC. Their staff, such as MOs, MPWs, and MPHS, follow RNTCP guidelines and are represented as 'roles', forming key users of the DSS shown in Figure 4(1).

The RNTCP hierarchy includes the Joint Secretary (administrative head), DDG-TB (program head), and CMO (CTD units). Each state has a State TB Officer (STO) leading the State TB Cell, and each district has a District TB Centre (DTC) led by a DTO, supported by SA and MO. TUs, headed by MO-TC and assisted by STS and STLS, coordinate multiple PHIs, both public and private. PHIs also include receiving and referring units for patient relocation, Directly Observed Treatment, Short-course (DOTS) Centers for treatment, and Sputum Collection Centers in remote areas, shown in Figure 4(1).

RNTCP comprises various processes carried out by different actors (continuants) across multiple locations (sites/spatial regions). As it evolves and depends on MoHFW, it is modeled as a 'process', specifically a subclass of 'history', shown in Figure 4(3). Nationwide activities such as training plans, policies, reviews, and action strategies are treated as occurrents under the RNTCP framework, unfolding over time and linked to corresponding continuants as shown in Figure 6.

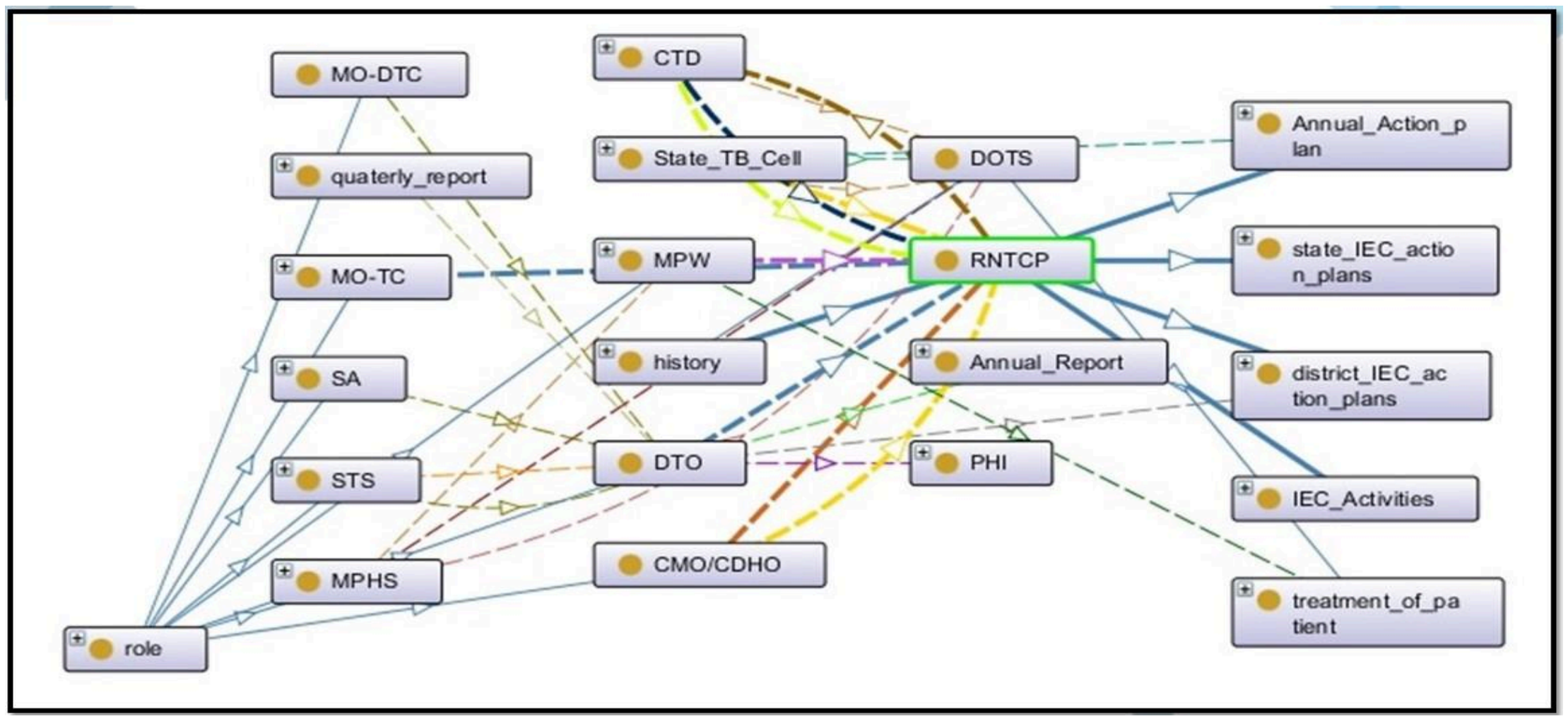

Figure 6. OntoGraf view of RNTCP and its classes

Within the RNTCP in TB ontology, various activities carried out by actors are categorized under occurrents, specifically as process boundaries, spatiotemporal regions, or temporal regions. A process boundary refers to a time-bound component of a larger process that lacks further temporal subdivisions, while a spatiotemporal region denotes an event or activity occurring within a defined space and time. Some processes fall under temporal regions, representing durations or points in time referenced against a standard timeline shown in Figure 4(3).

DOTS, a central component of RNTCP, is modeled as a 'process' that encapsulates various sub-processes. These include initial diagnosis procedures such as Sputum Smear Examination and treatment/medication, both of which are considered 'process profiles'. Despite serving different functions, the test and treatment are semantically connected, and treatment decisions are contingent upon diagnostic outcomes. Therefore, while the two processes are distinct, their realization within DOTS illustrates dependency and sequential progression.

This ontology-based classification clearly differentiates roles, actions, and events within TB management. By defining object properties (i.e., hasSymptom, prescribedBy, monitoredBy) and data properties (i.e., hasTemperature, hasWeightLoss, hasCoughDuration), the ontology captures the semantic relationships between patients, symptoms, healthcare providers, and treatments. These properties allow for rich semantic querying and rule-based reasoning. The model also reflects the complexity and interrelatedness of procedures in real-world healthcare workflows. Notably, even though DOTS is considered an effective cure, tuberculosis can manifest in different forms over a patient's lifetime, requiring repeated engagement with these interconnected processes and semantic entities.

### 3.5.2 Object and Data Properties of TB Ontology

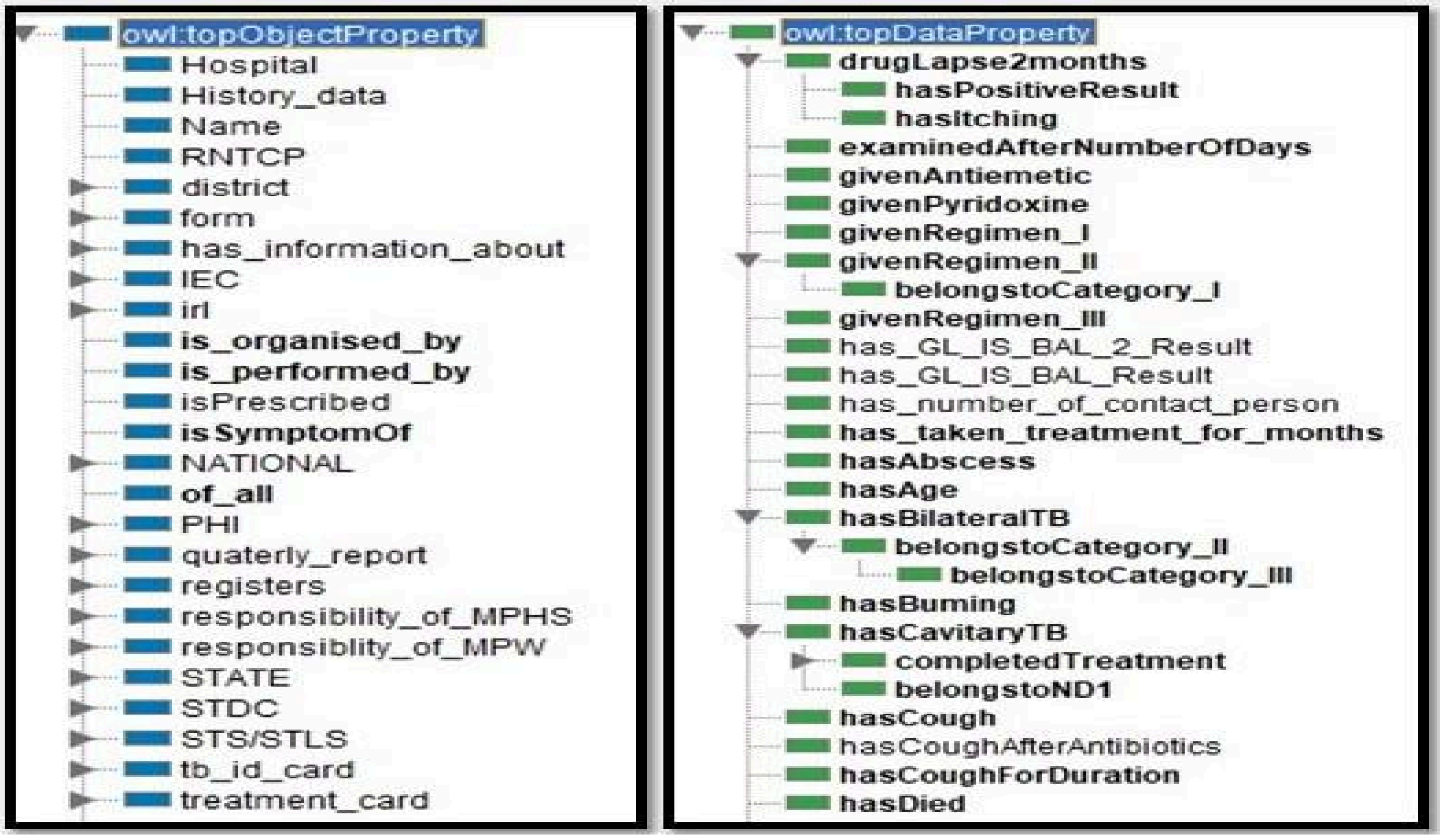

Figure 7. Object and Data Properties of TB ontology

In the ontology developed for TB management, data properties are used to define measurable or descriptive attributes of individual instances, such as symptoms, test results, and treatment status. Each data property is defined with a domain, representing the class it applies to, like *Patient*, and a range, indicating the type of value it holds, like *string*, *integer*, *boolean, and so on*. For instance, the data property *hasCoughDuration* may have a domain of *Patient* and a range of *integer*. All data properties are sub-properties of the general owl:topDataProperty developed TB ontology properties are shown in Figure 7.

Object properties, in contrast, define relationships between two class instances. In the TB domain, these are used to model interactions such as diagnosis, treatment, or administrative assignments. Each object property also has a specified domain and range, and is a sub-property of owl:topObjectProperty.

Examples:

- In the statement "Medical Officer treats Patient", the domain is *Medical Officer*, the range is *Patient*, and the object property *treats* expresses the connection.

- For administrative roles: "MoHFW appoints DTO", the domain is *MoHFW*, the range is *District TB Officer (DTO)*, and *appoints* is the object property linking the two entities.

These semantic properties ensure structured, machine-understandable relationships that support rule-based reasoning and decision-making in the TB Decision Support System. It is fundamental in enabling automated inference about patient status, recommended actions, and administrative workflows within the RNTCP framework.

**3.6 SWRL Rules Development**

The SWRL is integrated into the TB ontology to capture rule-based clinical logic derived from the RNTCP guidelines. These rules enable automated reasoning to support diagnosis and treatment recommendations for TB patients. For instance, symptoms such as persistent cough, fever, or lymph node swelling can activate SWRL rules that infer whether a sputum smear test is needed or identify the appropriate treatment category. Each rule is structured with an antecedent (the condition) and a consequent (the inferred action), allowing the system to reason over patient data and recommend actions accordingly.

These rules reflect the logic in RNTCP's decision-making flowcharts, ensuring alignment with standardized national protocols. Their semantic integration enables real-time, explainable, and uniform decision support. It is especially beneficial to field-level health workers such as ASHA and Medical Officers, who require timely and accurate guidance in managing TB cases.

Table 3 presents a set of SWRL rules that automatically classify TB patients into distinct disease stages based on clinical indicators and diagnostic results [29]. For example, in Stage 1 (Suspected TB), a patient with a cough lasting 14 days or more and fever is classified as *Suspected TB*. If sputum results are positive, the rule in Stage 2 categorizes the patient as having *Confirmed Pulmonary TB*. In Stage 3, patients with negative sputum results but lymph node enlargement over 2 cm are diagnosed with *Extra-Pulmonary TB*. Stage 4 (Severe TB) applies to patients with breathing difficulties, severe weight loss, and high-risk factors, while Stage 5 (Recovery Stage) identifies patients under DOTS therapy, with negative sputum and improved symptoms, as recovering.

These rules not only enable automated stage classification but also directly support treatment planning by aligning with RNTCP-prescribed regimens. As a result, the ontology-driven system enhances consistency, accuracy, and responsiveness in TB care, empowering health workers with structured and intelligent clinical decision support.

Table 3. SWRL rules for detecting the stage of TB disease

| Stage | Condition (SWRL Rule) | Action / Classification |
|---|---|---|
| Stage 1: Suspected TB | Patient(?p) ^ has_Cough_Duration(?p, ?d) ^ swrlb:greaterThanOrEqual(?d, 14) ^ has_Fever_Status(?p, "Yes") -> Suspected_TB(?p) | Classify the patient as Suspected TB |
| Stage 2: Confirmed Pulmonary TB | Patient(?p) ^ has_Sputum_Positive(?p, "Yes") -> Confirmed_Pulmonary_TB(?p) | Classify as Confirmed Pulmonary TB |
| Stage 3: Extra-Pulmonary TB | Patient(?p) ^ has_Sputum_Positive(?p, "No") ^ has_Lymph_Enlargement_Value(?p, ?v) ^ swrlb:greaterThan(?v, 2) -> Extra_Pulmonary_TB(?p) | Classify as Extra Pulmonary TB |
| Stage 4: Severe TB (Critical) | Patient(?p) ^ has_Breathing_Difficulty(?p, "Yes") ^ has_Weight_Loss(?p, "Severe") ^ has_Risk_Level(?p, "High") -> Severe_TB(?p) | Classify as SevereTB |
| Stage 5: Recovery Stage | Patient(?p) ^ is_Under_DOTS(?p, "Yes") ^ has_Sputum_Positive(?p, "No") ^ has_Symptom_Improvement(?p, "Yes") -> Recovery_Stage_TB(?p) | Classify as RecoveryStageTB |

Table 4 presents an enhanced set of SWRL rules designed to classify patients as suspected TB cases and recommend appropriate actions based on clinical symptoms, diagnostic findings, comorbid conditions, and epidemiological risk factors. These rules extend beyond traditional smear and X-ray criteria to include factors such as HIV status, diabetes, recent exposure history, high-prevalence area residency, and vulnerable populations like children and prison inmates. Each rule encodes expert knowledge into machine-interpretable logic, enabling automated TB risk assessment, early diagnosis, and timely initiation of treatment or preventive measures.

Table 4. Rules for classifying the patient as suspected TB

| **Serial No.** | **Rule for Condition** | **Action** | **SWRL Rule** |
|---|---|---|---|
| 1 | Persistent cough ≥ 3 weeks or coughing blood | Immediate sputum test and isolation | Patient(?p) ∧ has_Cough_For_Duration(?p, ?week) ∧ swrlb:greaterThanOrEqual(?week, 3) ∨ has_Haemoptysis(?p, "Yes") → undergoes(?p, sputum_test) ∧ isolation(?p, true) |
| 2 | Fever for ≥ 14 days with unexplained cause | Order TB screening | Patient(?p) ∧ has_Fever_Duration(?p, ?days) ∧ swrlb:greaterThanOrEqual(?days, 14) ∧ cause_Unexplained(?p, true) → undergoes(?p, tb_screening) |
| 3 | Close contact with a confirmed TB patient in last 6 months | Initiate preventive therapy | Patient(?p) ∧ contact_with_TB_Patient(?p, "Yes") ∧ contact_Period_Months(?p, ?m) ∧ swrlb:lessThanOrEqual(?m, 6) → given_Preventive_Therapy(?p, true) |
| 4 | HIV-positive patient with TB symptoms | Priority diagnostic testing | Patient(?p) ∧ has_HIV_Status(?p, "Positive") ∧ shows_TB_Symptoms(?p, true) → prioritize_Diagnostics(?p, true) |
| 5 | Positive Mantoux/TST test with abnormal chest X-ray | Classify as probable TB | Patient(?p) ∧ mantoux_Test_Result(?p, "Positive") ∧ has_Chest_Xray_Finding(?p, "Abnormal") → probable_TB(?p, true) |
| 6 | Weight loss > 10% in last 3 months | Flag for TB evaluation | Patient(?p) ∧ weight_Loss_Percentage(?p, ?w) ∧ swrlb:greaterThan(?w, 10) → tb_Evaluation_Required(?p, true) |
| 7 | Lymph node swelling in neck > 2 cm | Suggest TB lymphadenitis check | Patient(?p) ∧ lymph_Node_Swelling_Size(?p, ?cm) ∧ swrlb:greaterThan(?cm, 2) → check_TB_Lymphadenitis(?p, true) |
| 8 | Patient under 5 years old with TB exposure | Start prophylaxis | Patient(?p) ∧ age_Years(?p, ?a) ∧ swrlb:lessThan(?a, 5) ∧ contact_with_TB_Patient(?p, "Yes") → given_Prophylaxis(?p, true) |
| 9 | Chest X-ray shows cavities | Treat as active TB | Patient(?p) ∧ has_Chest_Xray_Finding(?p, "Cavities") → active_TB_Diagnosis(?p, true) |
| 10 | Past incomplete TB treatment | Start retreatment protocol | Patient(?p) ∧ treatment_History(?p, "Incomplete") → start_Retreatment_Protocol(?p, true) |
| 11 | Diabetic patient with TB symptoms | Accelerate testing | Patient(?p) ∧ has_Diabetes(?p, true) ∧ shows_TB_Symptoms(?p, true) → expedite_TB_Testing(?p, true) |
| 12 | Prison inmate with persistent cough | Conduct TB screening | Patient(?p) ∧ is_Prison_Inmate(?p, true) ∧ has_Cough_For_Duration(?p, ?week) ∧ swrlb:greaterThanOrEqual(?week, 2) → tb_Screening(?p, true) |
| 13 | Migrant from high TB prevalence area with symptoms | Flag for TB risk assessment | Patient(?p) ∧ from_High_TB_Prevalence_Area(?p, true) ∧ shows_TB_Symptoms(?p, true) → tb_Risk_Assessment(?p, true) |

Table 5 presents concise SWRL rules for identifying and managing pulmonary TB cases. The rules cover the diagnostic flow from initial symptoms (i.e, cough ≥ 2 weeks) to testing (sputum, X-ray), and classification (i.e, sputum-positive, high-risk, relapse). Treatment recommendations follow RNTCP guidelines, such as assigning Category I, starting Regimen I, or prioritizing high-risk patients. The rules also address relapse detection and latent TB risk through patient history and follow-up results. This rule-based approach supports automated, standardized, and early TB management decisions.

Table 5. SWRL rules for Confirmed Pulmonary TB patients

| **Serial No.** | **Rule for Condition** | **Action** | **SWRL Rule** |
|---|---|---|---|
| 1 | Patient with cough ≥ 2 weeks | Undergoes sputum test | Patient(?p) ∧ has_Cough(?p, ?value) ∧ has_Cough_For_Duration(?p, ?week) ∧ swrlb:equal(?value, "yes") ∧ swrlb:greaterThan(?week, 2) → undergoes(?p, sputum_1) |
| 2 | Patient with both smear results negative | Prescribed antibiotics for 14 days | antibiotics(?p) ∧ Patient(?p) ∧ has11_Smear_Result(?p, ?v) ∧ has12_SmearResult(?x, ?v1) ∧ swrlb:equal(?v, "negative") ∧ swrlb:equal(?v1, |
| | | | "negative") → is_Prescribed(?p, ?a) ∧ is_Prescribed_For_Duration(?p, 14) |
| 3 | Repeat smears negative | Undergo X-Ray Chest (XRC) and prescribe XRC | undergoes_Again(?p, ?s) ∧ has21_Smear_Result(?p, ?v) ∧ has22_Smear_Result(?p, ?v1) ∧ swrlb:equal(?v, "negative") ∧ swrlb:equal(?v1, "negative") → is_Prescribed(?p, xrc) ∧ undergoes(?p, xrc) |
| 4 | First smear positive | Mark as sputum-positive PTB | Patient(?p) ∧ has11_Smear_Result(?p, ?v1) ∧ swrlb:equal(?v1, "positive") → has_Sputum_Positive_PTB(?p, true) |
| 5 | Sputum-positive patient | Assign Category-I | Patient(?p) ∧ is_Smear_Positive(?p, true) → belongsto_Category_I(?p, true) |
| 6 | Patient belongs to Category-I | Start Regimen-I treatment | Patient(?p) ∧ belongsto_Category_I(?p, true) → given_Regimen_I(?p, true) |
| 7 | Cured patient reporting TB symptoms again | Retest sputum | Patient(?p) ∧ completed_Treatment(?p, true) ∧ is_Cured(?p, true) ∧ reports_Back_With_Symptom(?p, true) → undergoes_Again(?p, sputum_1) |
| 8 | Follow-up smear result 1 positive | Mark as a relapse | Patient(?p) ∧ undergoes_Again(?p, sputum_1) ∧ has31_Smear_Result(?p, ?value) ∧ swrlb:equal(?value, "positive") → is_Relapse(?p, true) |
| 9 | Follow-up smear result 2 positive | Confirm relapse | Patient(?p) ∧ undergoes_Again(?p, sputum_1) ∧ has32_Smear_Result(?p, ?value) ∧ swrlb:equal(?value, "positive") → is_Relapse(?p, true) |
| 10 | Negative sputum but abnormal chest X-ray | Confirm pulmonary TB | Patient(?p) ∧ has_Chest_Xray_Finding(?p, "Abnormal") ∧ has_Sputum_Positive(?p, "No") → Confirmed_PulmonaryTB(?p) |
| 11 | Severe weight loss and night sweats | Identify as a high-risk PTB patient | Patient(?p) ∧ has_Weight_Loss(?p, "Severe") ∧ has_Night_Sweats(?p, "Yes") → is_High_Risk_PTB(?p, true) |
| 12 | Patient marked as high-risk | Prioritize treatment | Patient(?p) ∧ is_High_Risk_PTB(?p, true) → prioritize_Treatment(?p, true) |
| 13 | The patient has a close contact history | Identified as latent TB at risk | Patient(?p) ∧ has_Contact_History(?p, "Yes") → is_Latent_TB_At_Risk(?p, true) |

Figure 8 presents a stage-by-stage model for diagnosing Extra-Pulmonary TB using SWRL rules within a semantic ontology framework. Specific clinical conditions define each diagnostic stage, and when these conditions are met, the corresponding SWRL rules are triggered. An LLM is integrated to generate clear, natural language explanations for each stage, making it easier for health workers to understand and act upon the diagnosis [24].

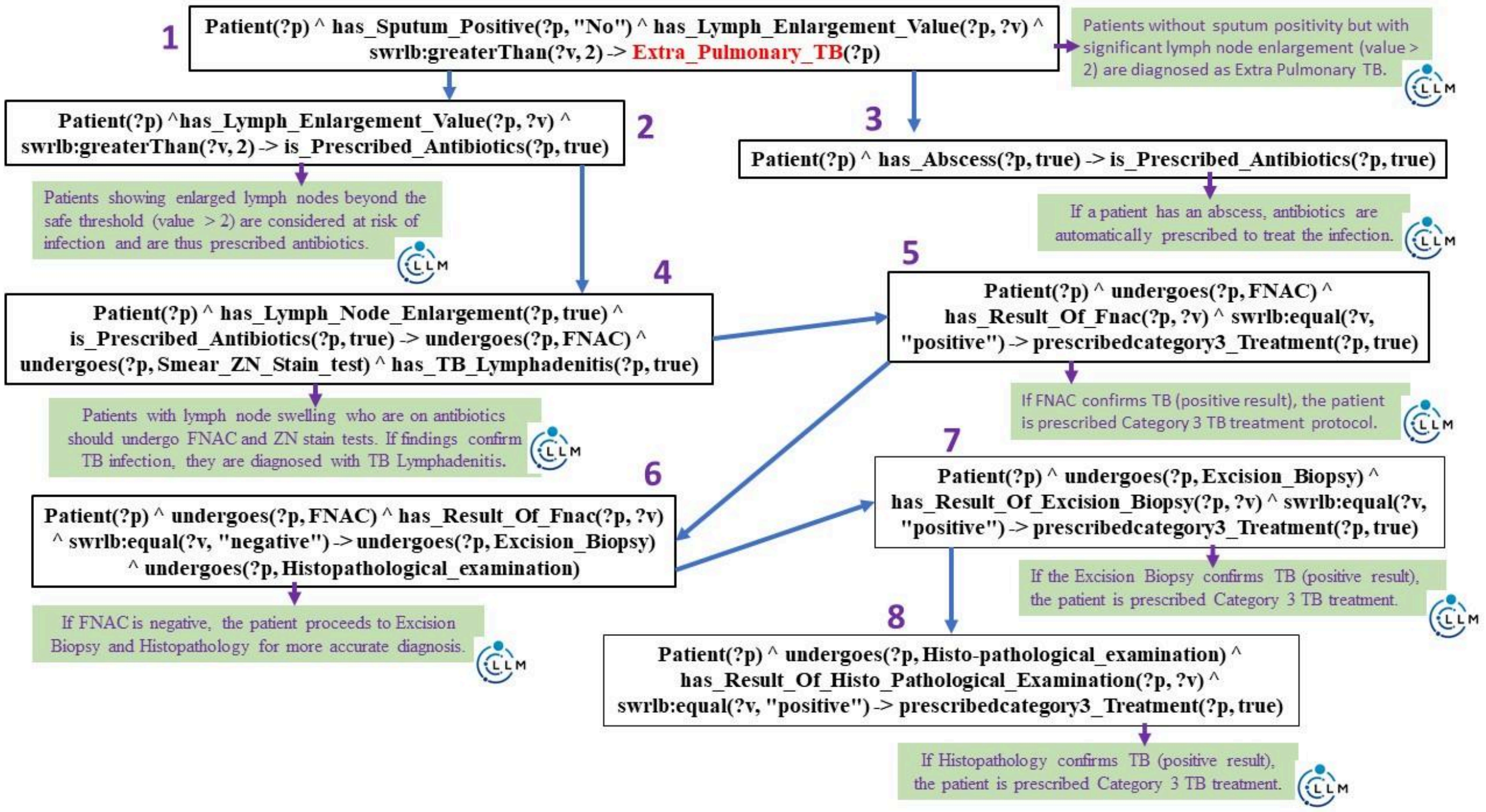

Figure 8. SWRL-based decision rules for Extra-Pulmonary TB patients with LLM-based explanation

Table 6 presents SWRL rules for diagnosing Severe TB by capturing critical clinical patterns. The process begins by identifying patients with breathing difficulty, severe weight loss, and high risk (Rule 1), and escalates based on findings like miliary TB on X-ray (Rule 2) or HIV co-infection (Rule 3). Children with altered mental status and TB meningitis symptoms are flagged as severe pediatric TB cases (Rule 4), while lung cavitations with high respiratory rate indicate severe pulmonary TB (Rule 5). ICU admission with multi-organ failure or signs of sepsis (Rules 6–7) signals critical TB conditions. Further rules detect extensive lung involvement (Rule 8), spinal TB with complications (Rule 9), and cardiac symptoms with pericardial effusion (Rule 10), supporting timely and structured diagnosis of severe TB forms.

Table 6. Rules for Severe TB diagnosis

| **Serial No.** | **Rule for Condition** | **Action (Diagnosis Step)** | **SWRL Rule** |
|---|---|---|---|
| 1 | Patient reports breathing difficulty, severe weight loss, and high risk | Flag as Potential Severe TB | Patient(?p) ∧ has_Breathing_Difficulty(?p, "Yes") ∧ has_Weight_Loss(?p, "Severe") ∧ has_Risk_Level(?p, "High") → is_Potential_Severe_TB(?p, true) |
| 2 | Radiological evidence of Miliary TB in potential severe TB case | Escalate to Radiologically Severe TB | is_Potential_Severe_TB(?p, true) ∧ has_Miliary_TB_Findings(?p, "Yes") → is_Radiological_Severe_TB(?p, true) |
| 3 | TB patient with HIV co-infection | Mark as Immunocompromised Severe TB | Patient(?p) ∧ has_TB_Co_Infection(?p, "HIV") → is_Immuno_Severe_TB(?p, true) |
| 4 | Pediatric patient with altered mental status and TB meningitis symptoms | Suspect Pediatric Meningeal TB | Patient(?p) ∧ is_Child(?p, "Yes") ∧ has_TB_Meningitis_Symptoms(?p, "Yes") ∧ has_Consciousness__Level(?p, "Altered") → is_Severe_Pediatric_TBM(?p, true) |
| 5 | Cavitary lesion on chest X-ray and respiratory rate > 30 | Suspect Severe Pulmonary TB | Patient(?p) ∧ has_Cavitary_Lesion(?p, "Yes") ∧ has_Respiratory_Rate(?p, ?r) ∧ swrlb:greaterThan(?r, 30) → is_Severe_Pulmonary_TB(?p, true) |
| 6 | TB patient in ICU with multi-organ failure | Flag as Critical TB Case | is_Suspected_TB(?p, true) ∧ is_Admitted_To_ICU(?p, "Yes") ∧ has_Multi_Organ_Failure(?p, "Yes") → is_Critical_TB(?p, true) |
| 7 | Confirmed TB case with sepsis indicators (i.e., hypotension, fever) | Suspect TB with Sepsis | Patient(?p) ∧ has_TB_Confirmed(?p, "Yes") ∧ has_Sepsis_Indicators(?p, "Yes") → is_Severe_Septic_TB(?p, true) |
| 8 | TB with extensive bilateral lung involvement on chest X-ray | Identify as Extensive Severe Pulmonary TB | Patient(?p) ∧ has_Bilateral_Involvement(?p, "Yes") ∧ has_Chest_Xray_Finding(?p, "Extensive") → is_Extensive_Pulm_TB(?p, true) |
| 9 | Spinal TB with neurological deficit (i.e., limb weakness) | Identify as Spinal TB with Complications | Patient(?p) ∧ has_Spinal_TB(?p, "Yes") ∧ has_Neurological_Deficit(?p, "Yes") → is_Complicated_Spinal_TB(?p, true) |
| 10 | Cardiac symptoms with pericardial effusion and confirmed TB | Flag as Severe TB with Cardiac Involvement | Patient(?p) ∧ has_Cardiac_Symptoms(?p, "Yes") ∧ has_Pericardial_Effusion(?p, "Yes") ∧ has_TB_Confirmed(?p, "Yes") → is_Cardiac_Severe_TB(?p, true) |

### 3.7 Kafka-Spark Streaming Pipeline for Real-Time Data Processing

After preprocessing, the TB dataset is ingested as real-time event streams using Apache Kafka's publish-subscribe model. The event producer gathers the cleaned clinical and patient data and publishes it to specific Kafka topics. These topics are partitioned and distributed across multiple brokers, Broker A, B, and C, each handling data replication, load balancing, and event coordination [30].

To ensure reliability and fault tolerance, Kafka replicates the event data across these brokers, safeguarding against data loss and ensuring high availability even if a broker fails. Once the data is securely replicated, it is streamed to Apache Spark Streaming for real-time analytics.

In the Spark Streaming layer, TB data is processed in micro-batches, as illustrated in Figure 9, with the implementation view presented in Figure 10. This setup enables continuous analysis, allowing for real-time pattern detection, event correlation, and timely updates—ultimately supporting rapid, data-driven decisions in TB diagnosis and treatment workflows.

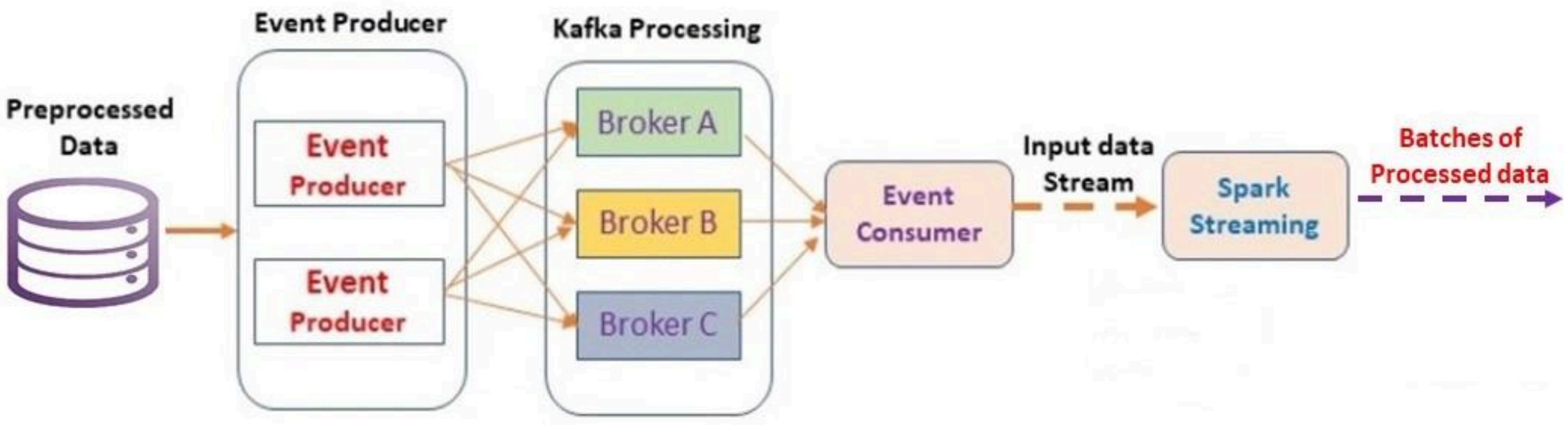

Figure 9. Kafka-Spark Streaming for Real-Time Data Processing

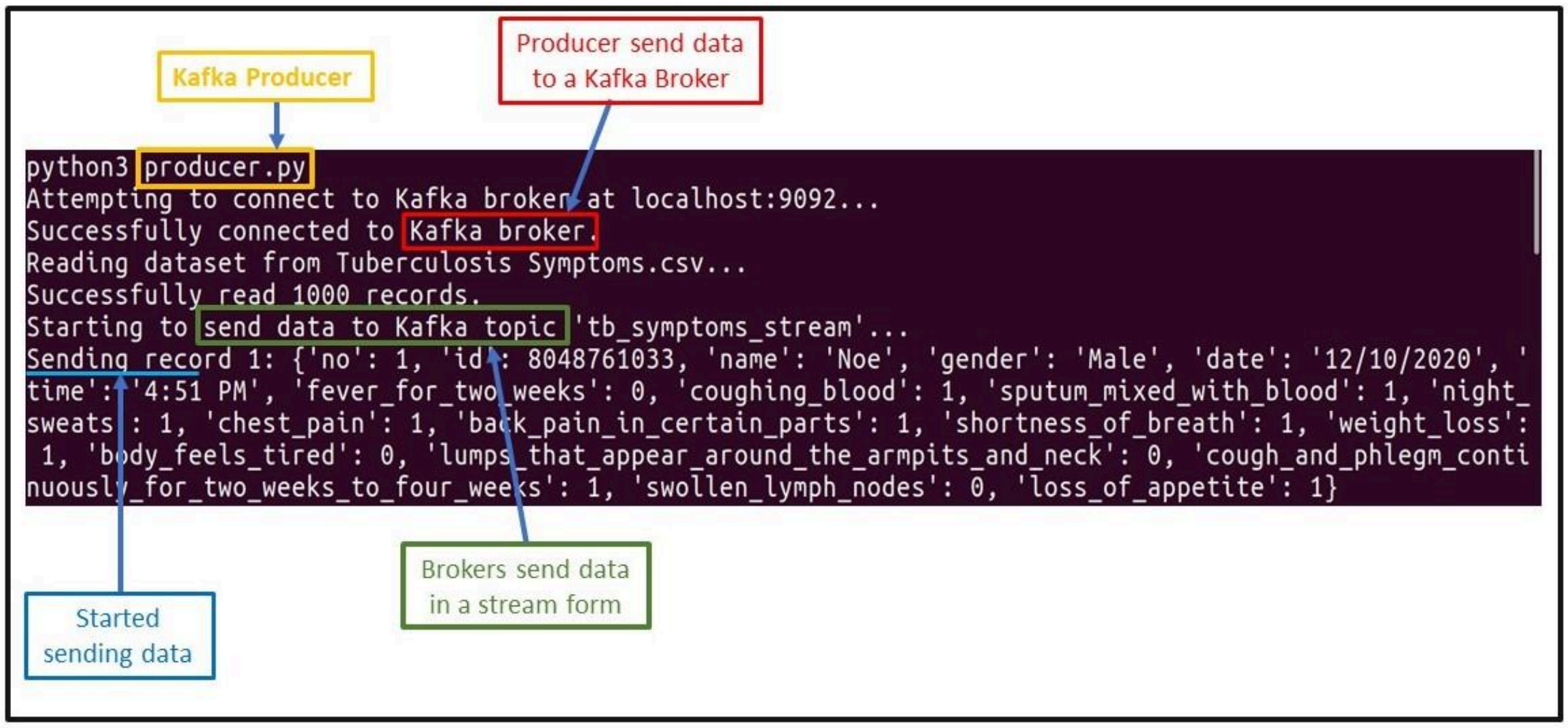

Figure 10. Implementation view of Kafka-Spark Streaming for Real-Time Data Processing

### 3.8 Complex Event Processing using Siddhi Engine

We utilize the Siddhi CEP engine to ingest real-time events from Apache Spark Streaming, enabling continuous monitoring and analysis of TB-related data. Siddhi CEP processes these data streams using predefined rule sets derived from standardized tuberculosis diagnosis and management protocols, allowing for the detection of complex clinical patterns [31].

As an open-source event processing engine, Siddhi is particularly well-suited for healthcare applications due to its capability to identify real-time correlations, patterns, and anomalies. In our framework, each event represents critical clinical indicators such as missed medication doses, abnormal laboratory results, or symptom escalation associated with TB patient monitoring.

Siddhi CEP applies rule-based logic to these incoming clinical parameters, estimating patient risk levels and detecting conditions that warrant medical intervention. When a complex event is identified, such as treatment non-adherence or symptom progression, the output is forwarded to two primary endpoints. First, it is sent to LLMs, which provide advanced clinical interpretation, generate context-aware recommendations, and deliver personalized insights to support adaptive decision-making. Second, the detected events are communicated to healthcare professionals and TB management teams, enabling timely risk assessment and informed clinical action. This real-time flow of clinical intelligence is illustrated in Figure 11.

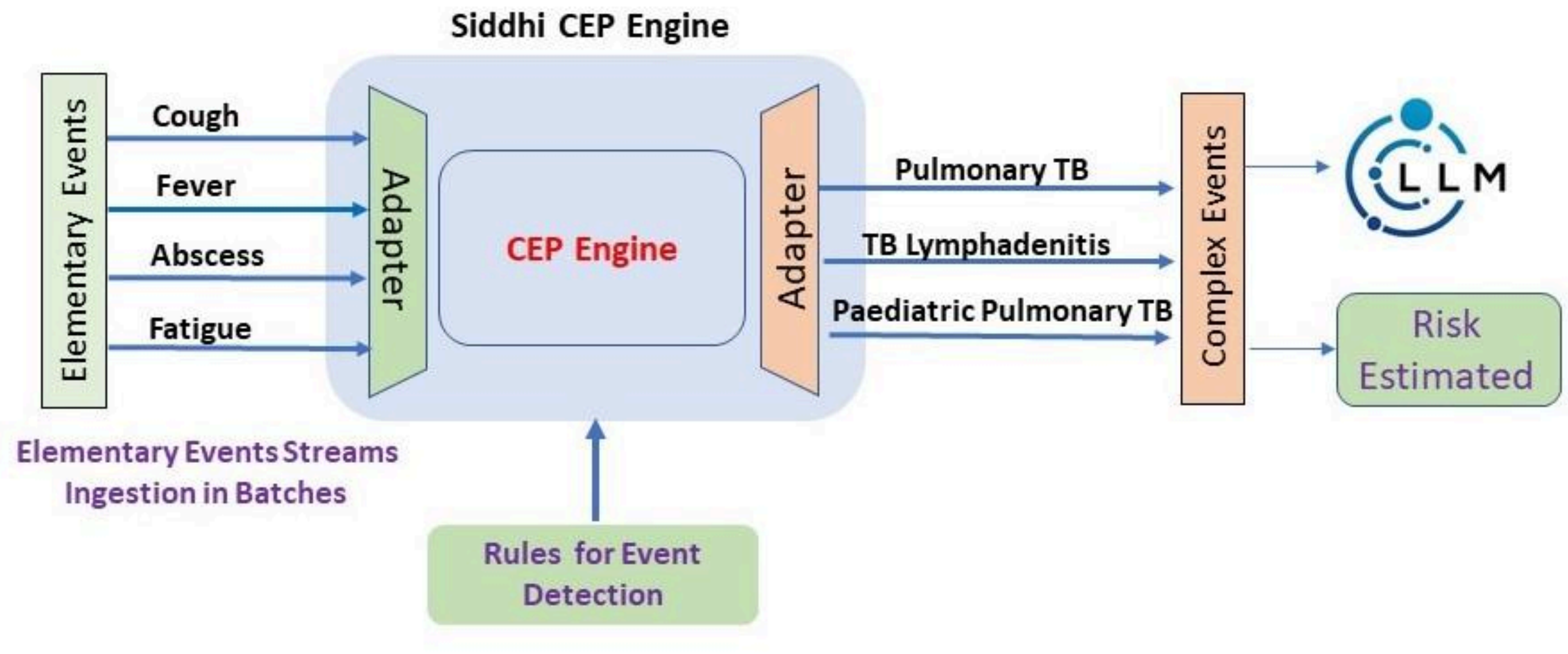

Figure 11. Siddhi CEP engine-based stream processing

Figure 12 shows the implementation view of Siddhi CEP for Alert generation based on the Rule.

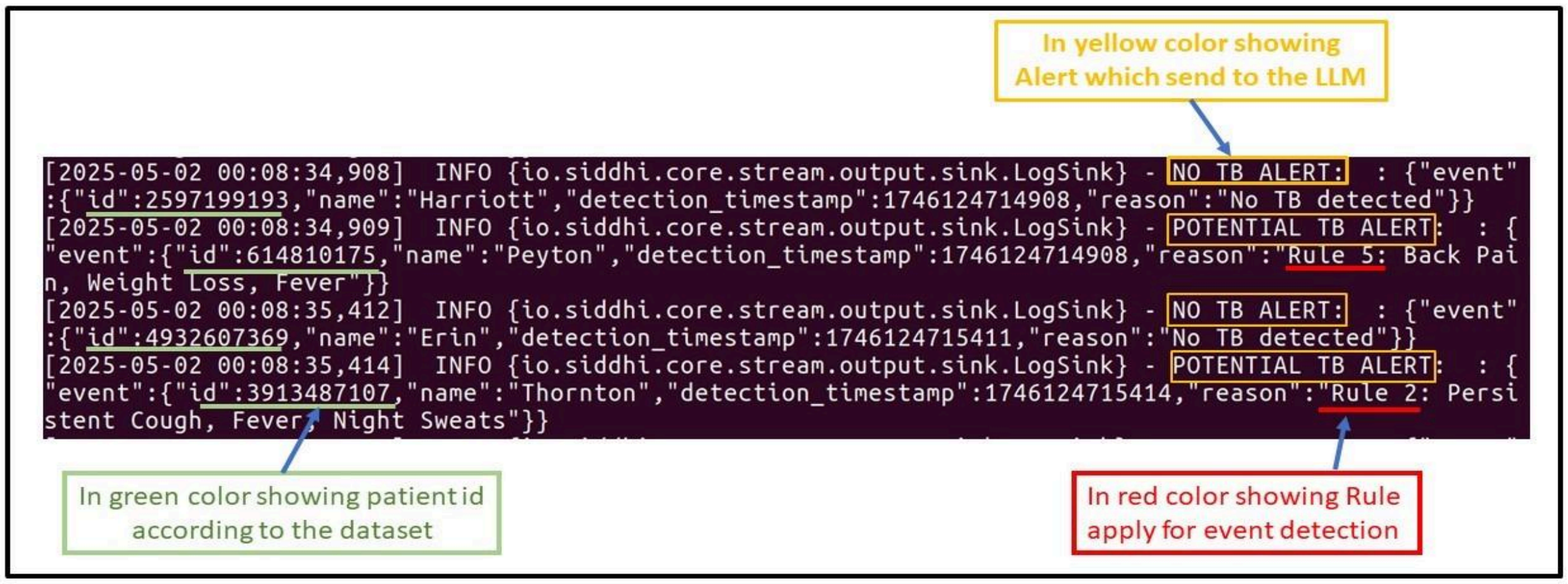

Figure 12. Implementation view of Siddhi CEP for Alert generation based on Rule

### 3.9 Complex Event-based Queries to LLMs for Clinical Interpretation

Advanced techniques are applied in this study to process and analyze TB clinical data, resulting in a structured and uniform textual format. The process initiates with a comprehensive preprocessing phase to eliminate errors, missing values, and noise to enhance data reliability and quality. Following this, text normalization ensures consistency across the dataset, preparing it for accurate downstream analysis.

To capture the semantic context of the data, normalized clinical guidelines are converted into document embeddings using the all-MiniLM-L6-v2[10] model. These embeddings encapsulate the underlying meaning of clinical narratives and patient information, allowing for precise and context-aware retrieval. Even subtle clinical indicators are preserved, supporting nuanced interpretation.

For efficient and rapid access, the embeddings are indexed using Facebook AI Similarity Search (FAISS) [32]. This indexing structure significantly accelerates response times to clinical queries, enabling healthcare systems to swiftly retrieve semantically relevant TB data for purposes such as diagnosis, monitoring, and decision support.

At the core of the system lies a question-and-answer (QA) module powered by the Intel/dynamic_tinybert[11] model, integrated within a robust retrieval framework. When a clinical query ranging from symptoms and diagnosis to treatment recommendations is submitted, the retriever searches the FAISS index to identify the two most contextually relevant documents. These retrieved materials are then passed to the QA model, which generates accurate, clinically relevant responses. To ensure continuity in ongoing interactions, the system incorporates a conversation buffer memory that preserves contextual flow across successive queries [33].

Integration with pre-trained LLMs and Apache Spark enhances the system's overall scalability, responsiveness, and adaptability. As illustrated in the architectural diagram in Figure 12, this design enables dynamic updates to align with evolving TB management protocols, new clinical guidelines, and changes in data formats, all while maintaining cost-effectiveness and reliability. The approach ensures real-time identification of critical TB cases, supporting timely intervention and better patient outcomes.

Real-time responsiveness is a key feature, empowering the system to process clinical queries rapidly and stream TB-related event data efficiently. Through the combined capabilities of Apache Spark and Siddhi CEP, the system handles large-scale datasets and manages numerous concurrent clinical queries with high throughput, making it especially effective in high-demand healthcare environments.

---

[10] https://huggingface.co/sentence-transformers/all-MiniLM-L6-v2
[11] https://huggingface.co/Intel/dynamic_tinybert

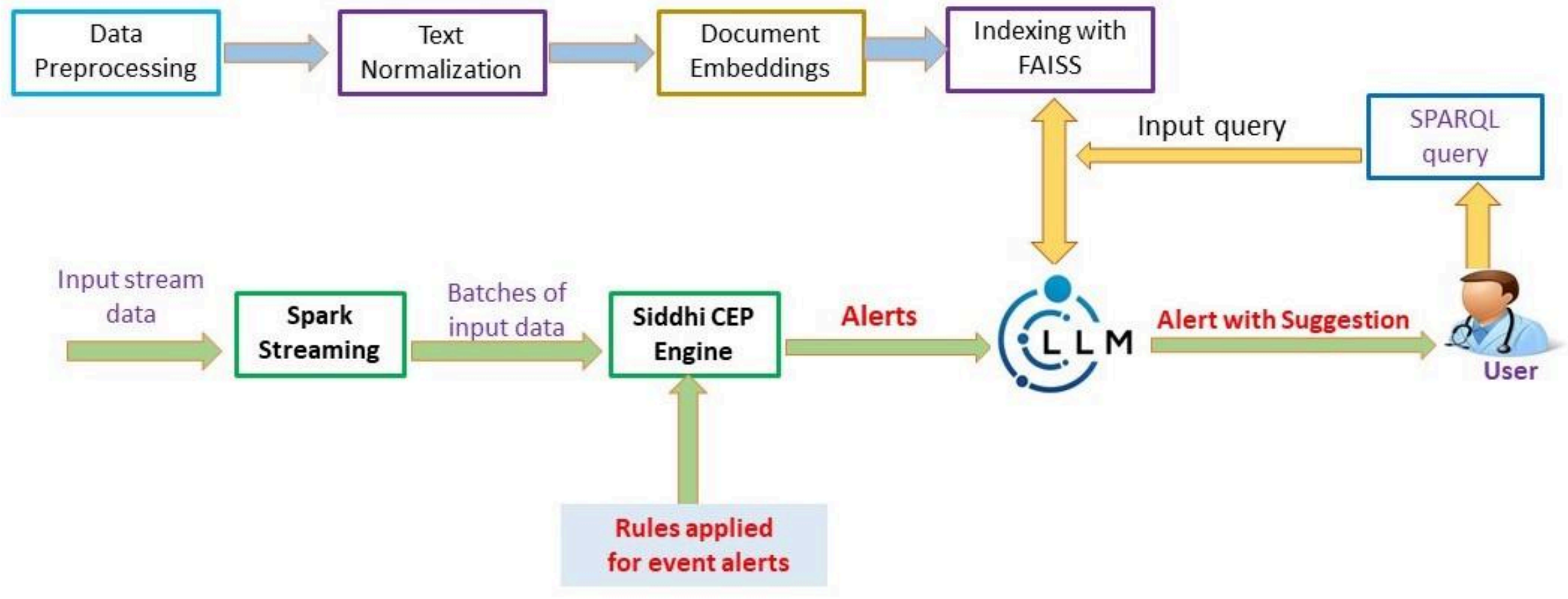

Figure 13**.** Integration of the Siddhi CEP Engine Framework and LLMs

Real-time responsiveness is seamlessly embedded within the system, enabling rapid processing of clinical queries and efficient transmission of TB-related event data. Leveraging Apache Spark and Siddhi CEP, the architecture is capable of managing vast TB datasets while handling multiple simultaneous clinical interactions without compromising performance. This design ensures robust scalability and maintains high throughput, making it particularly effective for deployment in resource-intensive and time-sensitive healthcare environments, as shown in Figure 13.

### 3.10 Generation of alerts via SPARQL queries on LLMs and TB ontology

When the same SPARQL query is run through the TB management framework, the LLM and ontology-based alert systems give answers that are very similar to each other [34]. But because it can learn more quickly, the LLM system gives more clinical information and personalized suggestions than just the results of a question. The ontology system checks the results that the LLM gives it using a structured knowledge base with diagnostic criteria, treatment standards, patient history, and clinical protocols. This double check makes sure that the LLM's suggestions are correct from a medical point of view and make sense in the context of TB. The joint process improves clinical decision support by making sure that AI-driven insights are accurate and in line with accepted standards for managing tuberculosis. Results are shown in Table 6.

Table 6. SPARQL Query results on TB Ontology and LLMs

| **SPARQL Query** |
|---|
| SELECT ?patient ?coughDuration ?feverStatus ?riskLevel<br>WHERE {<br>  ?patient rdf:type ex:TBPatient.<br>  ?patient ex:hasCoughDuration ?coughDuration.<br>  ?patient ex:hasFeverStatus ?feverStatus.<br>  ?patient ex:hasRiskLevel ?riskLevel. |

```
  FILTER (?coughDuration >= 14 && ?feverStatus = "Yes" && ?riskLevel = "High")
}
```

**TB Ontology-based Results**

| Patient ID | Cough Duration (days) | Fever Status | Risk Level |
|---|---|---|---|
| ex: Patient_109<br>ex: Patient _126<br>ex: Patient_164 | 18<br>21<br>16 | Yes<br>Yes<br>Yes | High<br>High<br>High |

**Predictive Insight (LLMs)**

Patient ex:PatientD, currently presenting with 12 days of persistent cough and intermittent fever, is projected to be classified as a high-risk TB suspect within the next 72 hours, based on clinical progression patterns and predicted symptom escalation.

**Precautionary Measures (LLMs)**

For ex:Patient_109, ex:Patient_126, and ex:Patient_164.

1. Schedule immediate sputum microscopy and chest X-ray.
2. Initiate contact tracing for household members.
3. Begin isolation protocols to prevent community transmission.

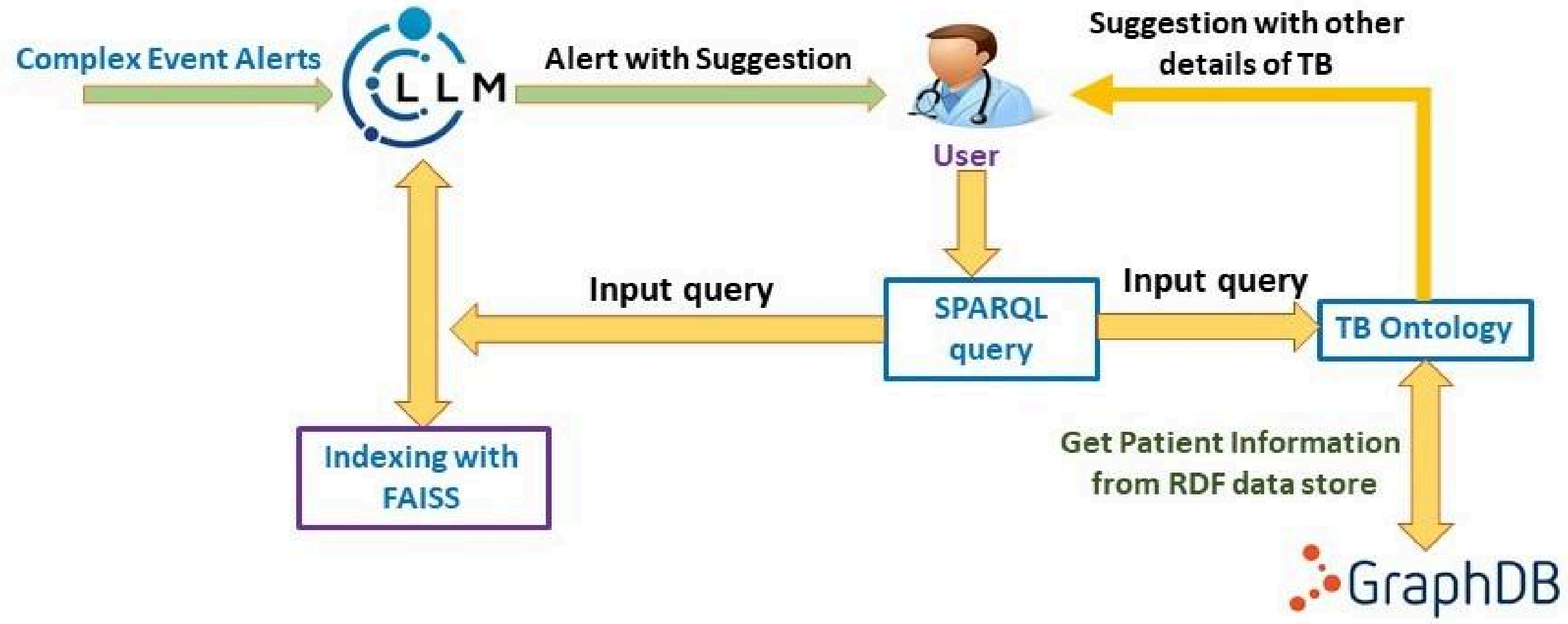

Figure 14. Integration of Siddhi CEP engine framework with LLMs, TB ontology, and RDF store in GraphD

Figure 14 highlights the critical role of the GraphDB RDF store in managing and querying semantically enriched tuberculosis data. GraphDB stores structured RDF triples derived from patient records, clinical observations, and ontology-based inferences as part of the integrated framework. These triples enable efficient SPARQL querying and semantic reasoning, supporting complex queries related to TB diagnosis, treatment history, and patient risk profiles. By linking the TB ontology with real-time data processed by Siddhi CEP and enriched by LLM outputs, GraphDB ensures that all clinical insights are grounded in a consistent, structured knowledge base. This allows healthcare professionals to perform precise searches, retrieve interconnected patient data, and derive actionable insights, enhancing the system's decision support capabilities.

### 3.11 LLMs-based TB ontology updates

This approach leverages LLMs to semi-automatically update the TB ontology by extracting new concepts, relationships, and rules from clinical documents such as guidelines and case records. The extracted knowledge is compared with the existing ontology to identify gaps, which are then reviewed and validated by domain experts [35]. Once approved, the ontology is updated and validated using Pellet reasoner in protégé to ensure consistency, and finally deployed in an RDF store like GraphDB for real-time semantic querying and decision support steps, as shown in Algorithm 1.

**Algorithm 1.** LLMs_Based_TB_Ontology_Update

**Input:** Clinical_Documents and Existing_TB_Ontology
**Output:** Updated_TB_Ontology

```
Begin

  Step 1: Data Collection
    Collect Clinical_Documents (guidelines, papers, patient data)
    Preprocess Documents (cleaning, tokenization)

  Step 2: Knowledge_Extraction
    For each Document in Clinical_Documents:
      Input Document into LLM
      Extract New_Concepts, New_Properties, New_Rules
      Store Extracted_Knowledge

  Step 3: Ontology_Update_Suggestion
    Compare Extracted_Knowledge with Existing_TB_Ontology
    Identify Missing_Classes, Missing_Properties, Missing_Rules
    Generate Suggested_Updates

  Step 4: Human_Validation
    Present Suggested_Updates to Domain_Experts
    For each Update in Suggested_Updates:
      If Expert_Approves(Update):
        Add Update to Validated_Updates

  Step 5: Apply Updates
```

```
        For each Update in Validated_Updates:
            If Update_Type == Class or Property:
                Add to TB_Ontology
            Else If Update_Type == Rule:
                Add SWRL_Rule to TB_Ontology

    Step 6: Ontology_Validation
        Run Reasoner on TB_Ontology
        If Ontology_Consistent:
            Save Updated_TB_Ontology

    Step 7: Deployment
        Store Updated_TB_Ontology in RDF_Store (e.g., GraphDB)

End
```

## 4. Result

This section presents the proposed system's evaluation, focusing on each key component's performance and quality. We assess the ontology's structural validity using standard schema-based metrics, analyze event processing efficiency, and evaluate the accuracy and reasoning capabilities of the LLM-integrated decision support system. The results demonstrate the framework's effectiveness in real-time TB detection, semantic enrichment, and intelligent decision-making. The following subsection details the ontology evaluation based on formal metrics that quantify its structural richness, consistency, and reasoning potential.

### 4.1 Ontology metrics schema-based evaluation

Based on the ontology metric counts presented in Figure 15, we derive several key structural metrics, including inheritance richness, relationship richness, class-to-relation ratio, and axiom-to-class ratio, as detailed in Table 7. An online tool[12] can be used to evaluate ontologies based on established metrics, results shown in Table 7.

[12] https://ontometrics.informatik.uni-rostock.de/ontologymetrics/

Ontology metrics:

**Metrics**

| | |
|---|---|
| Axiom | 3024 |
| Logical axiom count | 1866 |
| Declaration axioms count | 681 |
| Class count | 306 |
| Object property count | 193 |
| Data property count | 140 |
| Individual count | 18 |
| Annotation Property count | 26 |

**Class axioms**

| | |
|---|---|
| SubClassOf | 299 |
| EquivalentClasses | 8 |
| DisjointClasses | 668 |

Figure 15. TB Ontology metrics

Schema metrics describe an ontology using a 5-tuple model O = <C, Dr, Sc, Re, Ind>, where C represents the classes, Dr refers to data properties (attributes of classes), Sc defines subclass hierarchies, Re captures relationships between classes, and Ind denotes the individual instances within the ontology [29].

Various metrics are used to assess the structural complexity and quality of an ontology, including Attribute Richness (AR), which measures the average number of attributes per class; Relationship Richness (RR), which evaluates the diversity of relationships beyond class hierarchies; Class Richness (CR), which indicates the distribution of instances across classes; and Average Population (AP), which reflects the average number of instances per class.

RR measures the depth of connections between concepts in an ontology. It is calculated using Equation 1:

$$RR = \frac{|Prop|}{|Sub\ class| + |Prop|} \qquad \text{……………………………………..(1)}$$

where |Prop| is the total number of properties, including attribute data and object characteristics (class relationships).

AR is calculated by averaging the number of attributes over the entire class, as shown in Equation 2:

$$AR = \frac{|Attribute|}{|Class|} \qquad \text{….………………………………(2)}$$

where |attribute| represents the total number of data attributes.

CR indicates the amount of real-world knowledge conveyed through the ontology. It is calculated with equation 3 by dividing the number of classes with instances by the total number of classes:

$$CR = \frac{|\ Class\ with_instance\ |}{|Class|} \qquad \text{……………………………….(3)}$$

AP determines the average number of individuals in each class, expressed in equation 4:

$$AP = \frac{|Individual|}{|Class|} \quad \ldots\ldots\ldots\ldots\ldots\ldots\ldots\ldots\ldots\ldots\ldots\ldots (4)$$

The computed values of different evaluation metrics for the newly designed ontology are mentioned in Table 7.

Table 7: Different parameter-based metrics evaluation

| **Ontology Metrics** | **Results** |
|---|---|
| Attribute Richness | 0.496 |
| Class Richness | 0.017 |
| Average Population | 0.061 |
| Relationship Richness | 0.749 |

### 4.2 RDF Class Dependency and Ontology Rule Connectivity Analysis

Figure 15 extracted from GraphDB illustrates the RDF dependency structure of ontology classes, emphasizing the degree of interconnectivity among them. Core reasoning components such as

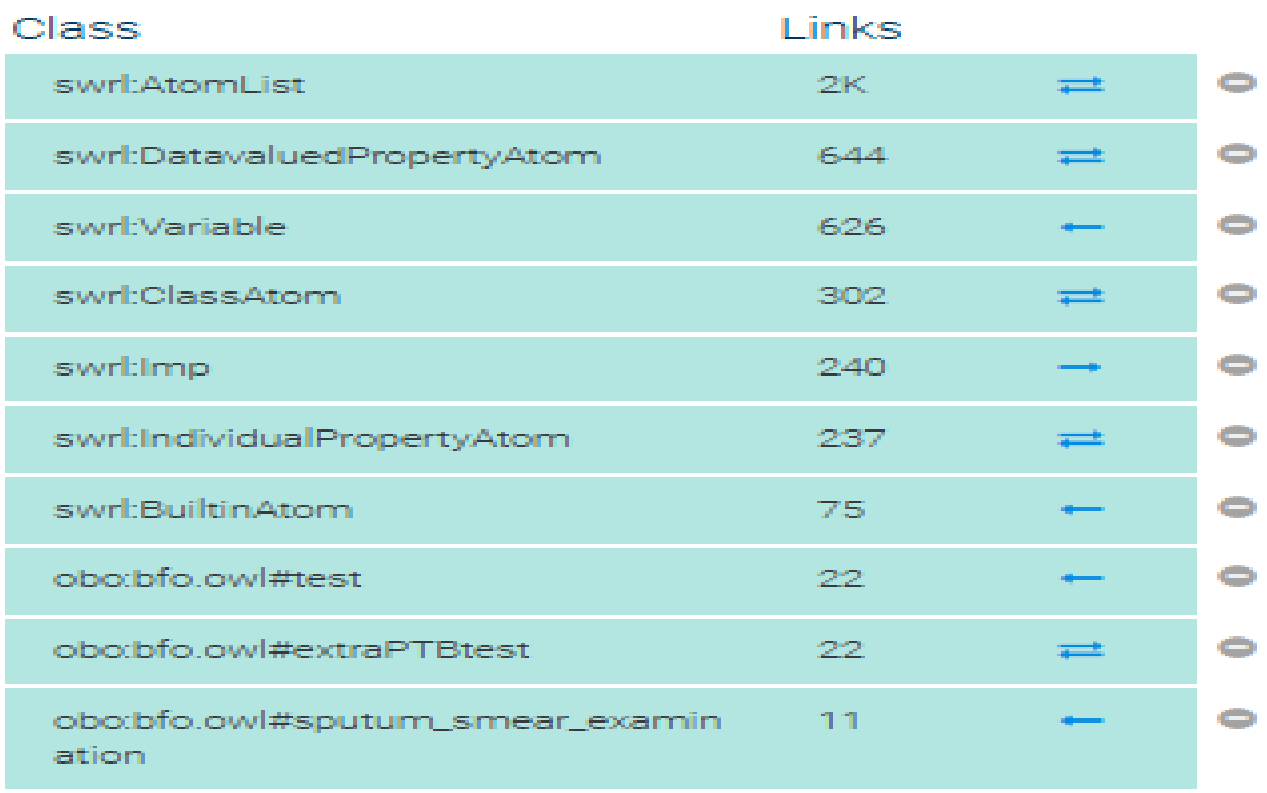

Figure 16. Shows Dependence classes and link in GraphDB[13]

swrl:AtomList, swrl:Variable, and swrl:DatavaluedPropertyAtom show high link density, reflecting extensive use of SWRL rules and logical constructs within the RDF model. These elements form the foundation for semantic inference, enabling automated rule execution across the ontology. In contrast, domain-specific RDF classes like *obo:bfo.owl#sputum_smear_examination* and *obo:bfo.owl#extraPTBtest* exhibit lower connectivity, indicating their roles as specialized concepts or leaf nodes. This visualization supports a structural evaluation of how RDF triples combine rule-based logic with domain-level semantics, facilitating efficient reasoning and semantic query processing. In detail of Figure 16 is illustrated in Table 8.

[13] https://sphn-semantic-framework.readthedocs.io/en/latest/user_guide/data_exploration.html

Table 8. RDF Class Dependencies Extracted from GraphDB within the Ontology

| Class | Links | Meaning |
|---|---|---|
| swrl:AtomList | 2K | Represents a list of SWRL atoms; heavily used in rule definitions, suggesting dense rule logic in the ontology. |
| swrl:DatavaluedPropertyAtom | 644 | Refers to SWRL atoms dealing with data property conditions (e.g., literals like numbers, strings). |
| swrl:Variable | 626 | Denotes variables used within SWRL rules, indicating a high number of rule-based operations. |
| swrl:ClassAtom | 302 | Specifies that class conditions are part of many rules, showing active usage of class constraints. |
| swrl:Imp | 240 | Represents SWRL implications (rules) — this number indicates how many rules are present. |
| swrl:IndividualPropertyAtom | 237 | Represents object property atoms connecting individuals, used in reasoning over relationships. |
| swrl:BuiltinAtom | 75 | Built-in functions used in rules (e.g., arithmetic, comparison). |
| obo:bfo.owl#test | 22 | A domain-specific class (perhaps a clinical test), with a small number of relations. |
| obo:bfo.owl#extraPTBtest | 22 | Likely a class related to pulmonary TB testing; moderately connected. |
| obo:bfo.owl#sputum_smear_examination | 11 | Represents a specific diagnostic test class; few direct connections, possibly used in specific rules. |

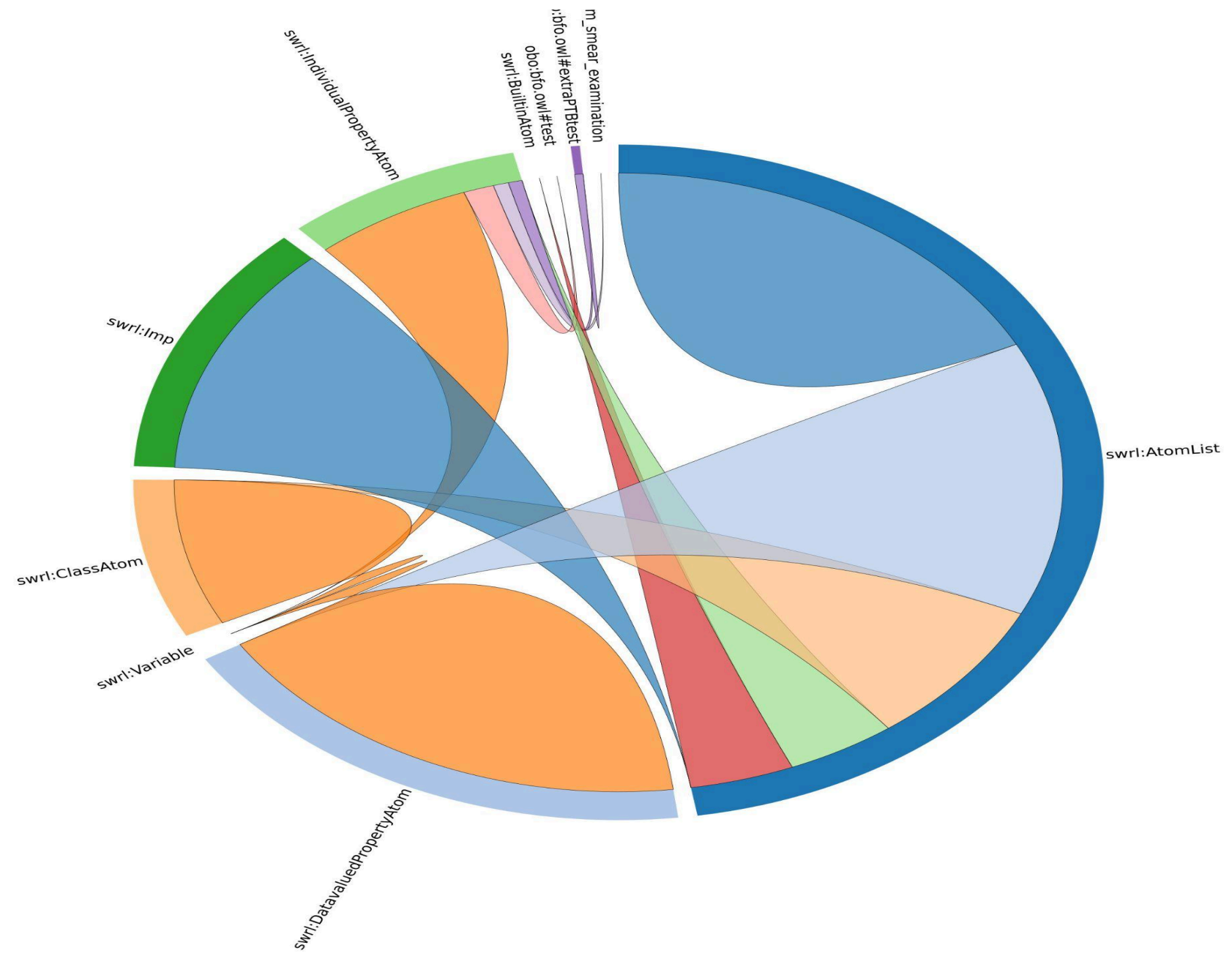

Figure 17. RDF Class Dependency Chord Diagram from GraphDB[14]

The chord diagram shown in Figure 17 offers a visual summary of how different RDF classes in the TB ontology are interlinked based on their usage and logical relationships within the dataset. Notably, core reasoning components such as *swrl:AtomList*, *swrl:Variable*, and *swrl:DatavaluedPropertyAtom dominate* the visualization with thick, numerous connections, indicating their frequent participation in rule definitions and inference logic. These components are the semantic backbone, supporting automated reasoning across various clinical scenarios. In contrast, domain-specific classes like *obo:bfo.owl#sputum_smear_examination* or *obo:bfo.owl#extraPTBtest* shows fewer connections, appearing as peripheral nodes or "leaf" entities. This pattern suggests that while the ontology is rich in clinical detail, its structural reasoning is concentrated around a core set of SWRL-based constructs, allowing efficient and focused semantic processing. The diagram illustrates how rule logic and domain concepts are structurally integrated to support scalable, knowledge-driven TB diagnosis.

### 4.3 Different SPARQL query results Analysis

Figure 18 shows the GraphDB framework view of how SPARQL query retrieves TB patients who have night sweats and fever for two weeks but no weight loss, indicating possible early-stage TB. It selects each patient's name, gender, and night sweats status, applying three filters in the WHERE clause to target patients with these early symptoms. By identifying individuals without severe progression, the query helps prioritize them for early medical intervention [36].

[14] https://graphdb.ontotext.com/documentation/11.0/visualize-and-explore.html

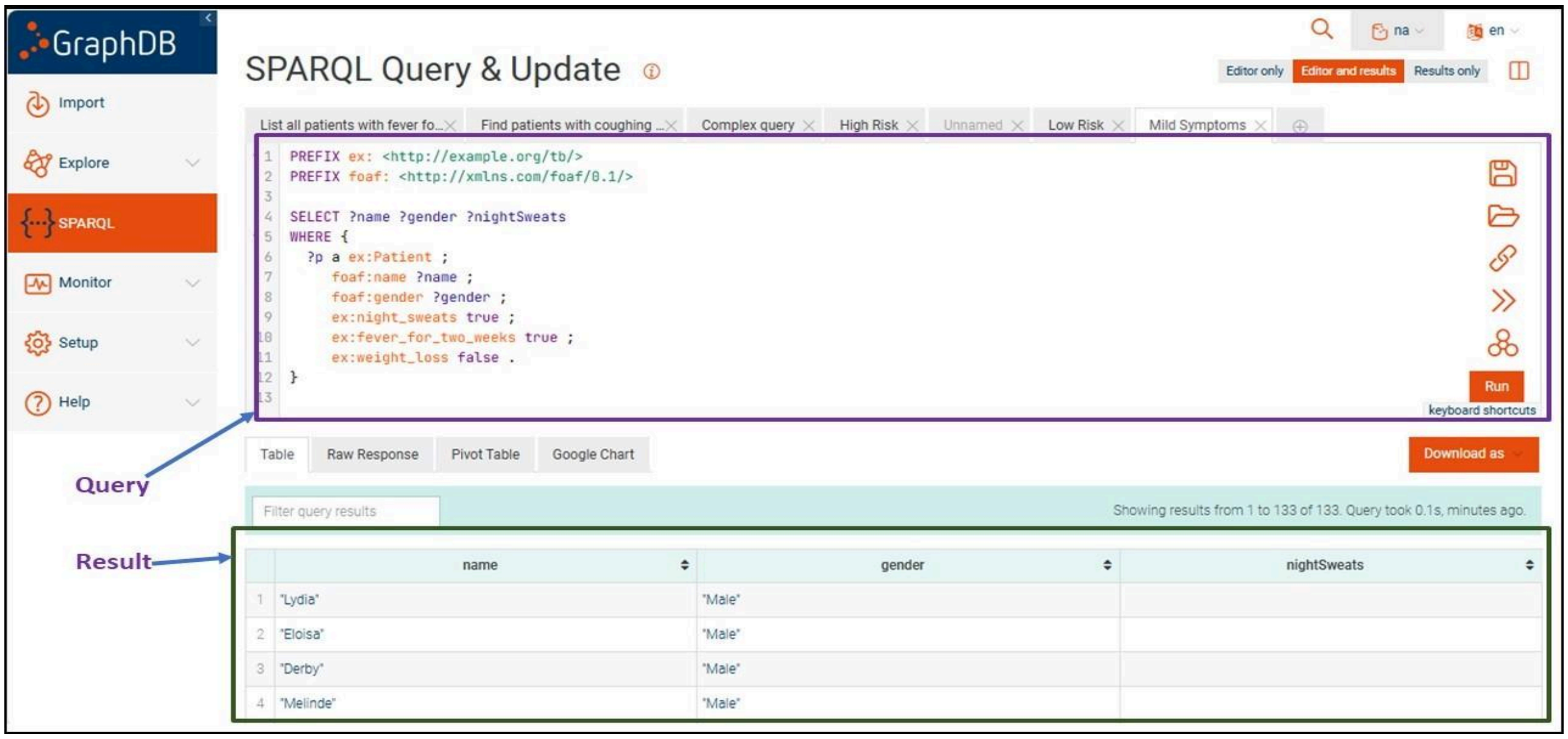

Figure 18. Framework view of GraphDB query interface

Table 9 presents a comparative analysis of seven SPARQL queries designed for TB patient data, detailing their query types, focus areas, key features, and execution times. Simple queries (0.20–0.25 seconds) are used for basic screenings, while theme-based and moderate queries (0.30–0.40 seconds) focus on early detection and symptom-specific monitoring. The complex query, incorporating temporal analysis, takes longer (0.90 seconds) and is intended for high-risk patient tracking over time.

Table 9. SPARQL Query Performance and Purpose Analysis for TB Patient RDF dataset

| Query No. | Query Theme / Purpose | Query Type | Time-Based? | Key Features Selected | Result Type | Time Taken in Education(Seconds) | Typical Use |
|---|---|---|---|---|---|---|---|
| 1 | Low-risk patients (no weight loss & no cough) | Simple | No | Name, gender | List of patients | 0.20 | Screening low-risk cases |
| 2 | Low-risk + fever check | Simple | No | Name, gender, fever status | List of patients | 0.25 | Early symptom monitoring |
| 3 | Early TB detection (fever + night sweats, no weight loss) | Theme-based | No | Name, gender, night sweats | List of early-stage cases | 0.30 | Monitor early TB signs |
| 4 | Females with cough but no blood | Theme-based | No | Name, cough status, sputum status | List of female patients | 0.35 | Focused screening |
| 5 | High-risk males (fever + weight loss) | Theme-based | No | Name, fever, weight loss | List of high-risk males | 0.30 | Prioritize for diagnosis |
| 6 | Swelling in neck & armpits (extrapulmonary TB risk) | Moderate Complexity | No | Name, lymph swelling, lumps | Swelling cases | 0.40 | Lymphatic TB detection |
| 7 | High-risk patients with symptom counts + date (complex) | Complex | Yes | Name, gender, date, symptom counts | High-risk + date | 0.90 | Risk assessment + trend tracking |

**4.4 Performance Evaluation of Rule-Based Event Processing Over Time Windows**

Figure 19 illustrates the event analysis conducted using a time window of 5 seconds to evaluate real-time processing of TB data. The graph highlights the system's ability to efficiently handle TB-related events within short time intervals. It is observed that as the window size increases, the number of processed events also grows, enhancing the effectiveness of rule-based diagnostic analysis. The evaluation was carried out by applying 5, 10, 15, 20, and 25 rules concurrently across varying window durations.

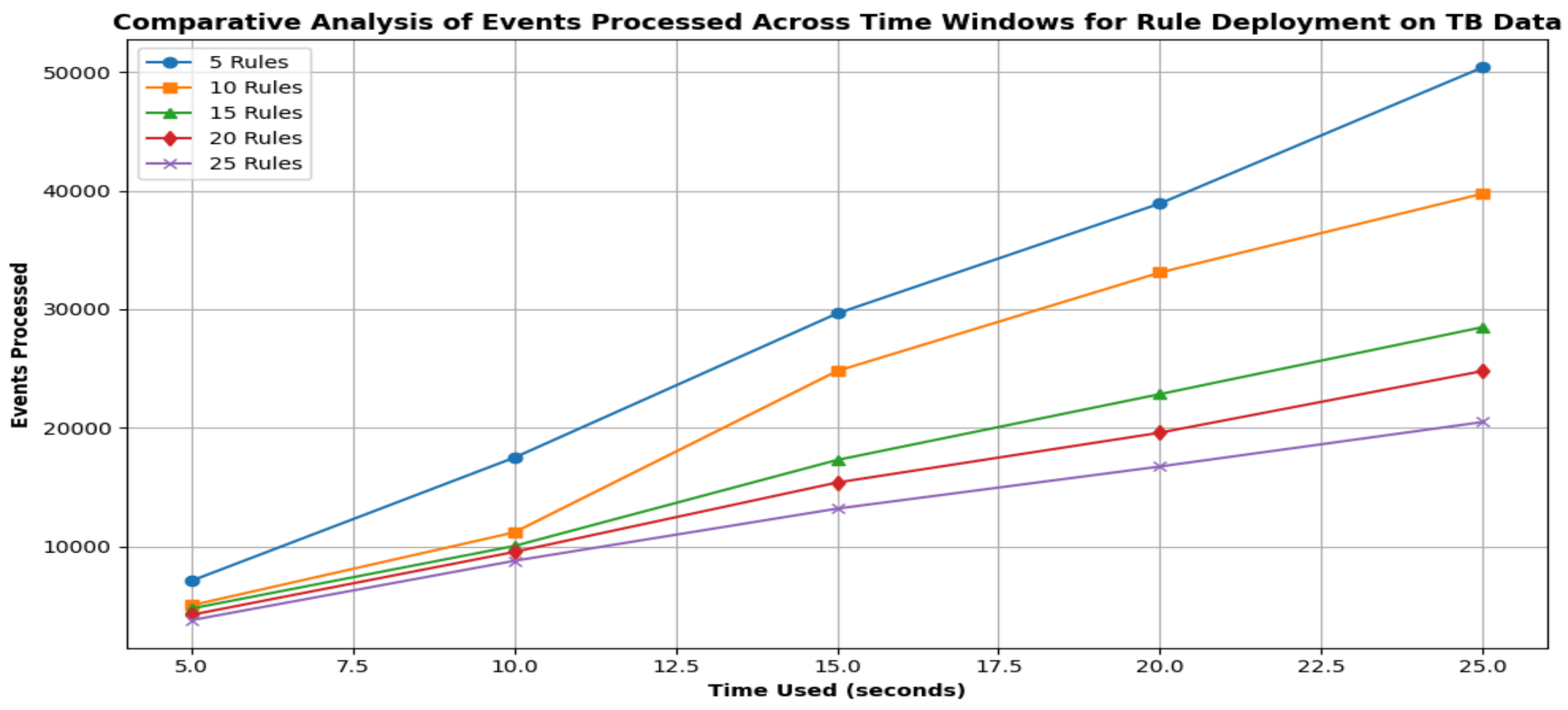

Figure 19. Deployment Time of Rules in Siddhi CEP Engine Across Varying Window Sizes

### 4.5 Query-based score Evaluation of LLM

To evaluate the performance of LLMs in TB diagnosis and knowledge retrieval, we first extract relevant clinical and ontology-based context from the FAISS database using the user's query. This context is ranked using cosine similarity, which measures the alignment between the query and the retrieved content values closer to +1 indicating higher relevance. The most relevant context is then input into the LLM, which generates a response based on this information. The quality of the generated response is evaluated using Precision, Recall, and F-measure by comparing it to a predefined, expert-validated reference answer. Precision indicates the proportion of medically relevant and accurate information in the response. Recall assesses how comprehensively the response covers the essential TB-related content, and the F-measure provides a balanced evaluation of both shown in Table 10. High metric values confirm that the LLM can deliver clinically relevant, context-aware responses, demonstrating its utility in TB diagnosis support and decision-making [37].

Table 10. Performance Score of LLMs on Query-Based Evaluation

| **Query** | **Similarity Score (FAISS)** | **Score of Response** | | |
|---|---|---|---|---|
| | | **Precision** | **Recall** | **F1-score** |
| What symptoms indicate early-stage TB without weight loss? | 0.8406 | 0.765 | 0.695 | 0.728 |
| What steps should be taken if a patient shows fever and night sweats but no cough or blood in sputum? | 0.9014 | 0.720 | 0.740 | 0.730 |
| How to identify patients with high-risk TB (persistent cough, weight loss, and swollen lymph nodes)? | 0.7633 | 0.815 | 0.745 | 0.778 |
| What symptoms cluster in patients with fever, no weight loss, and cough lasting two weeks? | 0.9879 | 0.690 | 0.715 | 0.702 |
| What are the main indicators of extrapulmonary TB (swollen lymph nodes and lumps in neck/armpit)? | 0.8109 | 0.770 | 0.750 | 0.760 |

Figure 20 shows the query result of Table 10.

```
question = "What symptoms indicate early-stage TB without weight loss?"
# Get the context chunks with scores
results = db.similarity_search_with_score(question)

# Sort by score (lower = better similarity)
best_doc, best_score = sorted(results, key=lambda x: x[1])[3]

# Print best matching chunk
print("Best Match (Similarity Score):", best_score)
print(best_doc.page_content)
```

```
Best Match (Similarity Score): 0.8406247
fever weight loss or night sweats are unlikely to have active tb and
should be offered preventive treatment regardless of art status in
addition to four symptom screening current cough fever weight loss or
night sweats cxr need to be done to exclude any abnormal radiological
findings suggestive of tb cxr is however not mandatory and lack of cxr
```

Figure 20. Implementation view of query 1 result of Table 10

Table 11 presents the results of the LLM evaluation in the context of TB, where the models operate without access to external clinical guidelines. Instead, they rely solely on the TB-specific dataset and related survey reports previously used in developing the TB ontology. Based on this setup, the outcomes include cosine similarity scores and the LLM-generated responses, categorized as true or false, after querying the models with TB-related questions.

Table 11. Query-Based Cosine Similarity and LLMs Response for TB Diagnosis (No False Cases)

| **Patient Query (Symptoms)** | **Cosine Similarity** | **LLMs Answer (Treatment Recommended?)** |
|---|---|---|
| Fever: Yes, Night Sweats: Yes, Weight Loss: No, Lymph Node Swelling: Yes, Cough: Yes, Blood in Sputum: No, Appetite Loss: No | 0.865 | True |
| Fever: Yes, Night Sweats: Yes, Weight Loss: Yes, Lymph Node Swelling: Yes, Persistent Cough: Yes, Blood in Sputum: Yes, Appetite Loss: Yes | 0.892 | True |
| Fever: Yes, Night Sweats: No, Weight Loss: No, Lymph Node Swelling: No, Persistent Cough: Yes, Blood in Sputum: No, Appetite Loss: No | 0.824 | True |
| Fever: No, Night Sweats: No, Weight Loss: No, Lymph Node Swelling: No, Persistent Cough: No, Blood in Sputum: No, Appetite Loss: No | 0.755 | True |
| Fever: Yes, Night Sweats: Yes, Weight Loss: Yes, Lymph Node Swelling: Yes, Persistent Cough: Yes, Blood in Sputum: Yes, Appetite Loss: Yes, Fatigue: Yes | 0.901 | True |

### 4.6 Model performance evaluation

Table 12 presents the diagnostic time distribution for five patients, illustrating the performance of each component in the proposed TB diagnosis system. The CEP module demonstrates consistently low execution times across all cases, enabling real-time event detection. In contrast, the LLM reasoning and ontology-based querying take comparatively more time, as they involve deeper semantic analysis and structured knowledge retrieval. Despite these additional steps, the overall execution time for the complete model combining CEP, LLM, and ontology is maintained within a reasonable range, ensuring both timely and accurate TB diagnosis.

Table 12. Execution time (in seconds) for CEP, LLM reasoning, ontology query, and the complete diagnosis model across five TB patients.

| Patient ID | CEP Time (s) | LLM Time (s) | Ontology Query Time (s) | Complete Model Time (s) |
|---|---|---|---|---|
| Patient 1 | 0.8 | 2.5 | 1.2 | 4.5 |
| Patient 2 | 0.9 | 2.3 | 1.1 | 4.3 |
| Patient 3 | 0.85 | 2.7 | 1.3 | 4.7 |
| Patient 4 | 0.75 | 2.6 | 1.0 | 4.2 |
| Patient 5 | 0.95 | 2.4 | 1.2 | 4.5 |

### 5. Discussion

This work addresses the challenge of TB detection in real-time big data environments by proposing a hybrid framework that integrates ontology-based reasoning, CEP, and LLMs. The system is evaluated across multiple components to ensure structural quality, reasoning efficiency, and responsiveness under streaming conditions.

The first stage evaluated the TB ontology using structural metrics, including class richness (the proportion of classes with instances, indicating conceptual coverage) and average path length (the average number of connections between ontology classes, reflecting semantic depth and navigability). These metrics helped assess the ontology's suitability for semantic reasoning, knowledge representation, and data integration.

To understand how domain-specific concepts and core reasoning components interact, we performed RDF class dependency analysis using GraphDB. This revealed the internal structure of the ontology, showing how SWRL constructs and TB-related concepts are interconnected. The results guided improvements in rule optimization and supported more efficient SPARQL query design.

SPARQL query performance was evaluated in terms of execution time for simple and complex queries within the GraphDB environment. The CEP engine, implemented using Siddhi, was tested based on how long it takes to apply rules across different window intervals (e.g., sliding and tumbling windows). We evaluated the LLM module's effectiveness using metrics such as similarity score, precision, recall, and F1 score on diagnostic queries.

We further analyzed the response time of each component ontology, CEP, and LLM, both independently and as an integrated system. Results show that the combined model balances semantic richness and real-time performance well.

A major challenge in scaling semantic systems for big data lies in the high overhead of ontology-based reasoning. The proposed TB ontology focuses only on essential diagnostic and symptomatic patterns to address this, making it lightweight and easier to integrate with stream processing tools like Apache Kafka and Siddhi CEP. Unlike generic ontologies, this domain-specific design avoids unnecessary reasoning overhead while capturing clinically relevant semantics [38].

CEP enables real-time monitoring of patient symptoms and event patterns. The ontology is used to define rules and guide event correlation, allowing for timely decision-making. Unlike traditional RDF systems that rely on heavy batch queries, the framework uses lightweight ontology lookups embedded in the event stream to maintain low latency.

RDF scalability is further improved by using GraphDB, a triplestore optimized for reasoning and indexing. Instead of performing continuous complex reasoning, the system triggers SPARQL queries selectively only when predefined SWRL-based event patterns occur. This strategy significantly reduces RDF query overhead and supports performance at scale.

To enhance adaptability, LLMs are integrated to suggest updates to the TB ontology by analyzing evolving symptom trends and medical literature. While the system does not support automated updates, expert-guided refinement based on LLM suggestions allows for semi-dynamic ontology evolution. This overcomes a key limitation of traditional semantic systems, which often remain static despite changes in medical knowledge or real-world data [38].

The system also addresses semantic reasoning bottlenecks by offloading high-complexity tasks to LLMs and reserving real-time tasks for lightweight SWRL rules within the CEP engine. This division ensures semantic expressiveness without compromising processing speed.

The proposed model was tested using a publicly available dataset of 1,000 TB patient records and demonstrated strong performance across all evaluation dimensions. However, the model has not yet been validated on private clinical datasets or in live hospital settings. A key limitation is that the ontology currently focuses solely on TB. As a result, the system may not accurately classify cases where TB symptoms overlap with other diseases or where co-morbidities exist. To address this, ontologies for related conditions should be integrated, along with new rule definitions to capture overlapping and multi-condition patterns. Although LLMs can suggest these extensions, domain experts must review and implement final updates manually.

### 5.1 Comparison with existing works

Integrating ontology-driven CEP with LLMs, the proposed system addresses critical gaps left by prior approaches. It overcomes challenges such as semantic heterogeneity, rigid rule structures, and limited adaptability by combining semantic reasoning, real-time data stream processing, and intelligent LLM-based decision-making. Unlike LLM-assisted rule generation [40] or fuzzy rule-based CEP models [17], it provides greater contextual awareness and flexibility. Ontology-based systems like [22] and [11] focus on semantic modeling or schema generation but lack the capacity for real-time, adaptive reasoning. By bridging structured ontological inference with unstructured data interpretation, the proposed framework delivers a more comprehensive, low-latency, and accurate solution for real-time tuberculosis surveillance and healthcare analytics, as shown in Table 13.

Table 13. Comparison based on existing related techniques

| Reference | Target Domain | Research Problem Addressed | Methodology Used | Result |
|---|---|---|---|---|
| [39] | Internet of Multimedia Things / CEP | Integration challenges between LLM-based multi-agent systems and CEP frameworks for dynamic event processing | Development of an LLM-based Multi-Agent System (MAS) using AutoGen + Kafka pub/sub for CEP; performance evaluation using video queries | Demonstrated functional LLM-MAS pipeline with high narrative coherence, though higher complexity increases latency |
| [22] | Healthcare / Big Data Analytics | Semantic interoperability and knowledge-driven reasoning in CEP for real-time healthcare data | Ontology-based CEP architecture (OCEP) using RDF, SPARQL, SSN/SOSA ontology, Kafka, and Hadoop | Achieved 85% accuracy in real-time healthcare event detection and improved decision support using PPG data |
| [40] | Federated CEP / Distributed Systems | Time-consuming and error-prone CEP rule generation relying on domain experts; lack of proactive rule generation | Federated LLM-assisted rule generation and refinement using prompt engineering (Zero-Shot, CoT, ToT, etc.) and distributed rule testing | Prompt engineering (Few-Shot + CoT) improved activity recognition; federated setup shows promise for CEP rule adaptability and accuracy |
| [11] | Primary Healthcare – Health Data Storage and Management | No standard method for designing NoSQL schemas tailored for heterogeneous, large-scale health data; poor query performance with relational models. | Proposed an ontology-driven NoSQL schema generation algorithm using ontology concepts, sample queries, their statistics, and performance constraints; implemented in MongoDB. | Significantly improved query response times compared to relational schema; effective handling of polymorphic health data types using ontology-driven design. |
| [17] | Cardiovascular Disease Prediction in IoT-enabled Health Systems | Difficulty in managing real-time disease prediction due to dynamic nature of cardiovascular parameters and limitations in traditional CEP rule design. | Developed a fuzzy rule-based CEP system using Apache Kafka, Apache Spark, and Siddhi CEP; fuzzy logic was used to define risk rules based on WHO standards; real-time data stream processing. | System categorized cardiovascular risk into 5 levels using fuzzy rules; validated with 1000 synthetic samples; achieved fast, real-time, accurate, and adaptive prediction outcomes. |
| [24] | Healthcare – Medical Text Mining | Challenges in extracting structured knowledge from unstructured clinical data due to lack of semantic understanding and domain-specific context. | Designed an ontology-based NLP framework that integrates domain ontology, rule-based parsing, named entity recognition, and semantic annotation to extract medical concepts and their relations. | Successfully extracted relevant healthcare entities and relations with improved precision and recall; facilitated better clinical data structuring for decision-support systems. |
| **Proposed** | Real-Time Tuberculosis Surveillance and Healthcare Analytics | Limitations of traditional CEP systems in handling semantic heterogeneity, unstructured data, and real-time reasoning for TB detection. | Developed an Ontology-based CEP framework integrated with LLMs. Used TB-specific ontology for semantic context, Siddhi CEP for event stream processing, Apache Kafka for ingestion, and Spark for distributed computation. LLMs were used for knowledge-based reasoning and anomaly detection. | Demonstrated improved event detection accuracy, low-latency decision-making, semantic enrichment of TB symptoms and case profiles, and better performance than conventional CEP systems. |

As shown in Table 14, existing TB-related research has primarily focused on ontology development, enrichment, and integration. For instance, [41] and [43] concentrated on creating structured ontologies for TB diagnosis and case modeling, yet lacked real-time processing capabilities. Efforts such as [42] employed semi-automatic text mining for ontology enrichment, while [25] addressed data fragmentation through ontology-based data access for improved querying. In contrast, the proposed approach integrates ontology-driven CEP with LLM reasoning, enabling dynamic analysis of unstructured health data streams. This combination allows for intelligent, low-latency tuberculosis event detection and semantic context awareness, delivering a more adaptive and scalable solution for real-time TB surveillance and decision support.

Table 14. Comparison with existing works related to TB diseases

| Reference | Target Domain | Research Problem Addressed | Methodology Used | Result |
|---|---|---|---|---|
| [41] | Pulmonary Tuberculosis – Public Health & Epidemiology | Lack of structured domain knowledge for supporting TB diagnosis and treatment tools. | Developed OntoTB using SABiO methodology and UFO foundational ontology; used OntoUML modeling; validated with domain experts. | Produced OntoTB reference ontology covering diagnosis, treatment, and prevention; enabled semantic organization of TB knowledge for decision support and data collection tools. |
| [42] | Tuberculosis in Epidemiology Domain | Manual ontology construction is time-consuming; need scalable enrichment for TB ontologies. | Semi-automatic text mining on 200 scientific articles; term extraction with POS tagging, Text2Onto, Dog4dag; validation by epidemiologists. | Generated enriched TB ontology with 121 concepts and 11 object properties; successfully merged with Epidemiology Ontology (EPO); ensured reasoning consistency. |
| [25] | TB Surveillance – Clinical and Epidemiological Integration | Fragmentation between clinical and epidemiological data in TB surveillance systems. | Developed integrated ontology-driven system architecture; used Protégé, SPARQL, Ontology-Based Data Access (OBDA) model for linking relational and RDF data. | Achieved improved TB surveillance via integrated querying; ontology-based views enabled knowledge discovery from heterogeneous data. |
| [43] | Tuberculosis Case Management | Inconsistent modeling of TB case data hampers decision support. | Designed OntoTBC, an OWL ontology integrating clinical concepts and case progression using Protégé and DL reasoning. | Provided structured, reusable TB case representation; enabled consistency checking and semantic querying for clinical decision support. |
| **Proposed** | Real-Time Tuberculosis Surveillance and Healthcare Analytics | Limitations of traditional CEP systems in handling semantic heterogeneity, unstructured data, and real-time reasoning for TB detection. | Developed an Ontology-based CEP framework integrated with LLMs. Used TB-specific ontology for semantic context, Siddhi CEP for event stream processing, Apache Kafka for ingestion, and Spark for distributed computation. LLMs were used for knowledge-based reasoning and anomaly detection. | Demonstrated improved event detection accuracy, low-latency decision-making, semantic enrichment of TB symptoms and case profiles, and better performance than conventional CEP systems. |

**6. Conclusion and Future Work**

In this study, we proposed an integrated framework for intelligent tuberculosis detection that leverages the strengths of ontologies, CEP, and LLMs within a Big Data ecosystem. By combining Apache Kafka for real-time data ingestion, Apache Spark for stream analytics, and the Siddhi CEP engine for pattern detection, the system facilitates high-throughput, low-latency processing of clinical event streams. The ontology component ensures semantic interoperability and contextual reasoning, while LLMs enhance interpretability and domain-specific inference over RDF data. Experimental results using real-world publicly available TB datasets demonstrated the model's capability to detect complex symptoms and risk patterns with high precision, offering a powerful decision-support tool for early diagnosis and intervention.

Despite the promising results, several challenges remain. The system uses a semi-automatic ontology update mechanism supported by LLMs; however, achieving fully autonomous updates in response to evolving clinical guidelines remains a complex task. Furthermore, integration with real-time hospital information systems and validation using private clinical datasets have not yet been implemented.

For future work, we aim to:

1. Extend the framework to support multi-disease detection by enriching the ontology with co-morbid conditions.
2. Incorporate online learning mechanisms to update LLM reasoning based on clinician feedback and new data patterns.
3. Develop a FHIR-compliant interface for seamless integration with Electronic Health Record (EHR) systems.
4. Deploy and evaluate the system in a real clinical setting to assess performance under real-world constraints, including patient privacy, interoperability, and decision latency.

**Acknowledgment**

The author gratefully for the support received under the National Fellowship for Persons with Disabilities (NFPWD) scheme from the University Grants Commission (UGC), India, for pursuing Ph.D. research. We also thank the Indian Institute of Information Technology, Allahabad, for providing the necessary infrastructure and resources. Special thanks to the Big Data Analytics (BDA) Lab members for their valuable input and support in this entire work.

**Declaration**

**Competing interests**

The authors declare no competing interests relevant to the content of this work.

**Author's contribution statement**

**Ritesh Chandra**: Conceptualization, Data curation, Methodology, Visualization, Writing – original draft. **Sonali Agarwal**: Formal analysis, Investigation, Supervision, Writing – review & editing. **Navjot Singh**: Supervision, Writing – review & editing

**Data availability and access**

The study utilized freely accessible data obtained from a website. We extend our thanks to the authors and collaborators for providing the original data.

**Funding**

No funding received for this work.

**Conflicts of interests**

All authors declare that they have no conflicts of interest in the presented work.

**Clinical trial number**

Not applicable.

**References**

[1] Sathiyamoorthi, S., Tiwari, U., Muralikrishnan, S., Aravindakshan, R., & Ganapathy, K. (2025). Impact of the COVID-19 Pandemic on Tuberculosis: A Retrospective Analytical Study of Morbidity Profiles, Trends, and Patient Care in a Primary Tuberculosis Treatment Unit in India. Cureus, 17(4).

[2] Malwe, S., Bawiskar, D., Wagh, V., & WAGH, V. (2023). Tuberculosis and the effectiveness of the revised National Tuberculosis Control Program (RNTCP) to control tuberculosis: a narrative review. Cureus, 15(12).
[3] Valentini, R. (2024). Ontology-based Data Management in Healthcare.
[4] Madani, K., Russo, C., & Rinaldi, A. M. (2019, December). Merging large ontologies using bigdata graphdb. In 2019 IEEE international conference on big data (big data) (pp. 2383-2392). IEEE.
[5] Daniel, G., Sunyé, G., & Cabot, J. (2016, October). UMLtoGraphDB: mapping conceptual schemas to graph databases. In International Conference on Conceptual Modeling (pp. 430-444). Cham: Springer International Publishing.
[6] Jung, K. H. (2025). Large language models in medicine: Clinical applications, technical challenges, and ethical considerations. Healthcare Informatics Research, 31(2), 114-124.
[7] Konys, A. (2016, October). Ontology-based approaches to big data analytics. In International multi-conference on advanced computer systems (pp. 355-365). Cham: Springer International Publishing.
[8] Ullah, F., Habib, M. A., Farhan, M., Khalid, S., Durrani, M. Y., & Jabbar, S. (2017). Semantic interoperability for big-data in heterogeneous IoT infrastructure for healthcare. Sustainable cities and society, 34, 90-96.
[9] Bharambe, U., Narvekar, C., & Andugula, P. (2022). Ontology and knowledge graphs for semantic analysis in natural language processing. In Graph Learning and Network Science for Natural Language Processing (pp. 105-130). CRC Press.
[10] Li, H., Hartig, O., Armiento, R., & Lambrix, P. (2024). Ontology-based GraphQL server generation for data access and data integration. Semantic Web, 15(5), 1639-1675.
[11] Sen, P. S., & Mukherjee, N. (2024). An ontology-based approach to designing a NoSQL database for semi-structured and unstructured health data. Cluster computing, 27(1), 959-976.
[12] Croce, F., Valentini, R., Maranghi, M., Grani, G., Lenzerini, M., & Rosati, R. (2024). Ontology-based data preparation in healthcare: The case of the AMD-STITCH project. SN Computer Science, 5(4), 437.
[13] Ahmed, H., Younis, E. M., Hendawi, A., & Ali, A. A. (2020). Heart disease identification from patients' social posts, machine learning solution on Spark. Future Generation Computer Systems, 111, 714-722.
[14] Hiraman, B. R. (2018, August). A study of apache kafka in big data stream processing. In 2018 International Conference on Information, Communication, Engineering and Technology (ICICET) (pp. 1-3). IEEE.
[15] Vennamaneni, P. R. (2025). Real-Time Financial Data Processing Using Apache Spark and Kafka. International journal of data science and machine learning, 5(01), 137-169.
[16] Rahmani, A. M., Babaei, Z., & Souri, A. (2021). Event-driven IoT architecture for data analysis of reliable healthcare application using complex event processing. Cluster Computing, 24(2), 1347-1360.
[17] Kumar, S. S., Chandra, R., Harsh, A., & Agarwal, S. (2025). Fuzzy rule-based intelligent cardiovascular disease prediction using complex event processing. The Journal of Supercomputing, 81(2), 402.
[18] Kuwar, V., Sonwaney, V., Upreti, S., Ekatpure, S. R., Divakaran, P., Upreti, K., & Poonia, R. C. (2025). Real-Time data analytics and decision making in Cyber-Physical systems. In Navigating Cyber-Physical Systems With Cutting-Edge Technologies (pp. 373-390). IGI Global Scientific Publishing.

[19] Mehandru, N., Miao, B. Y., Almaraz, E. R., Sushil, M., Butte, A. J., & Alaa, A. (2024). Evaluating large language models as agents in the clinic. NPJ digital medicine, 7(1), 84.
[20] Hager, P., Jungmann, F., Holland, R., Bhagat, K., Hubrecht, I., Knauer, M., ... & Rueckert, D. (2024). Evaluation and mitigation of the limitations of large language models in clinical decision-making. Nature medicine, 30(9), 2613-2622.
[21] Wang, Z., Li, H., Huang, D., Kim, H. S., Shin, C. W., & Rahmani, A. M. (2025). Healthq: Unveiling questioning capabilities of llm chains in healthcare conversations. Smart Health, 100570.
[22] Chandra, R., Agarwal, S., Kumar, S. S., & Singh, N. (2025). OCEP: An Ontology-Based Complex Event Processing Framework for Healthcare Decision Support in Big Data Analytics. arXiv preprint arXiv:2503.21453.
[23] Mavridis, A., Tegos, S., Anastasiou, C., Papoutsoglou, M., & Meditskos, G. (2025). Large language models for intelligent RDF knowledge graph construction: results from medical ontology mapping. Frontiers in Artificial Intelligence, 8, 1546179.
[24] Chandra, R., Tiwari, S., Rastogi, S., & Agarwal, S. (2025). A diagnosis and treatment of liver diseases: integrating batch processing, rule-based event detection and explainable artificial intelligence. Evolving Systems, 16(2), 1-26.
[25] Jiomekong, A., Tapamo, H., & Camara, G. (2023, October). An ontology for tuberculosis surveillance system. In Iberoamerica*n Knowledge Graphs and Semantic Web Conference* (pp. 1-15). Cham: Springer Nature Switzerland.
[26] Kokash, N., Wang, L., Gillespie, T. H., Belloum, A., Grosso, P., Quinney, S., ... & de Bono, B. (2025). Ontology-and LLM-based Data Harmonization for Federated Learning in Healthcare. arXiv preprint arXiv:2505.20020.
[27] Guarnier, T. S., da Silva Teixeira, M. D. G., & Costa, D. D. R. (2021). Tuberculosis diagnosis–An ontology-driven conceptual model. Proceedings http://ceur-ws. org ISSN, 1613, 0073.
[28] Otte, J. N., Beverley, J., & Ruttenberg, A. (2022). BFO: Basic formal ontology. Applied ontology, 17(1), 17-43.
[29] Chandra, R., Tiwari, S., Agarwal, S., & Singh, N. (2023). Semantic web-based diagnosis and treatment of vector-borne diseases using SWRL rules. Knowledge-Based Systems, 274, 110645.
[30] Lekkala, C. (2022). Integration of Real-Time Data Streaming Technologies in Hybrid Cloud Environments: Kafka, Spark, and Kubernetes. European Journal of Advances in Engineering and Technology, 9(10), 38-43.
[31] Kumar, S. S., Chandra, R., & Agarwal, S. (2024). A real-time approach for smart building operations prediction using rule-based complex event processing and SPARQL query. The Journal of Supercomputing, 80(15), 21569-21591.
[32] Krisnawati, L. D., Mahastama, A. W., Haw, S. C., Ng, K. W., & Naveen, P. (2024). Indonesian-English Textual Similarity Detection Using Universal Sentence Encoder (USE) and Facebook AI Similarity Search (FAISS). CommIT (Communication and Information Technology) Journal, 18(2), 183-195.
[33] Han, D., Lim, S., Roh, D., Kim, S., Kim, S., & Yi, M. Y. (2025, January). Leveraging LLM-Generated Schema Descriptions for Unanswerable Question Detection in Clinical Data. In Proceedings of the 31st International Conference on Computational Linguistics (pp. 10594-10601).
[34] Allemang, D., & Sequeda, J. (2024, November). Increasing the accuracy of LLM question-answering systems with ontologies. In International Semantic Web Conference (pp. 324-339). Cham: Springer Nature Switzerland.

[35] Liu, Z., Gan, C., Wang, J., Zhang, Y., Bo, Z., Sun, M., ... & Zhang, W. (2025, April). Ontotune: Ontology-driven self-training for aligning large language models. In Proceedings of the ACM on Web Conference 2025 (pp. 119-133).
[36] Kumar, S. S., Chandra, R., & Agarwal, S. (2024). Rule based complex event processing for an air quality monitoring system in smart city. Sustainable Cities and Society, 112, 105609.
[37] Xu, S., Wu, Z., Zhao, H., Shu, P., Liu, Z., Liao, W., ... & Li, X. (2024). Reasoning before comparison: LLM-enhanced semantic similarity metrics for domain specialized text analysis. arXiv preprint arXiv:2402.11398.
[38] Hoseini, S., Theissen-Lipp, J., & Quix, C. (2024). A survey on semantic data management as intersection of ontology-based data access, semantic modeling and data lakes. Journal of Web Semantics, 81, 100819.
[39] Zeeshan, T., Kumar, A., Pirttikangas, S., & Tarkoma, S. (2025). Large language model based multi-agent system augmented complex event processing pipeline for internet of multimedia things. arXiv preprint arXiv:2501.00906.
[40] Lotfian Delouee, M., Pernes, D. G., Degeler, V., & Koldehofe, B. (2024). Poster: Towards Federated LLM-Powered CEP Rule Generation and Refinement.
[41] Guarnier, T. S., da Silva Teixeira, M. D. G., & Costa, D. D. R. (2021). Tuberculosis diagnosis–An ontology-driven conceptual model. Proceedings http://ceur-ws. org ISSN, 1613, 0073.
[42] Ramayanti, Desi, et al. "Tuberculosis ontology generation and enrichment based text mining." 2020 international conference on information technology systems and innovation (ICITSI). IEEE, 2020.
[43] Ogundele, O. A., Moodley, D., Seebregts, C. J., & Pillay, A. W. (2015, September). An ontology for tuberculosis treatment adherence behaviour. In Proceedings of the 2015 Annual Research Conference on South African Institute of Computer Scientists and Information Technologists (pp. 1-10).